\documentclass[preprint,10pt,3p]{elsarticle}
\usepackage{graphics} 
\usepackage{epsfig} 
\usepackage{times} 
\usepackage{amsmath} 
\usepackage{amssymb}  
\usepackage{epstopdf}
\usepackage{subfigure,color,balance}
\usepackage{caption}
\usepackage{verbatim}
\usepackage{cases}
\allowdisplaybreaks[4]
\usepackage{enumerate}
\usepackage{booktabs}
\usepackage{balance}
\usepackage{mathbbol}
\usepackage{algorithm}
\usepackage{algorithmicx}
\usepackage{algpseudocode}
\usepackage{mathrsfs}
\usepackage{threeparttable}
\usepackage[draft]{todonotes}
\usepackage{makecell}

\usepackage{enumitem}

\newcommand{\method}[1]{\texttt{#1}}

\usepackage[colorlinks=true,      
linkcolor=black,      
citecolor=black,      
filecolor=black,      
urlcolor=blue]{hyperref}

\newtheorem{definition}{Definition}
\newtheorem{theorem}{Theorem}

\newdefinition{remark}{Remark}
\newtheorem{lemma}{Lemma}
\newtheorem{corollary}{Corollary}

\newtheorem{assumption}{Assumption}

\def\Y{{\bf Y}}

\def\x{{\bf x}}

\def\0{{\bf 0}}
\def\1{{\bf 1}}

\def\col{{\mathrm {col}}}
\def\st{{\mathrm {subject~to}}}

\def\eg{{\em e.g.}}
\def\ie{{\em i.e.}}

\def\diag{{\rm diag}}

\def\argmin{\mathop{\rm argmin}}

 \biboptions{compress}

\begin{document}
	
	\begin{frontmatter}
		
		\title{{Distributed data-driven predictive control for cooperatively smoothing \\ mixed traffic flow}} 
		
		
		
		\author[a,b]{Jiawei Wang}
		\ead{wang-jw18@mails.tsinghua.edu.cn}
		\author[b]{Yingzhao Lian}
		\ead{yingzhao.lian@epfl.ch}
		\author[b]{Yuning Jiang\corref{cor1}}
		\ead{yuning.jiang@epfl.ch}
		\author[a]{Qing Xu}
		\ead{qingxu@tsinghua.edu.cn}
		\author[a]{Keqiang Li}
		\ead{likq@tsinghua.edu.cn}
		\author[b]{Colin N. Jones}
		\ead{colin.jones@epfl.ch}
		
		\cortext[cor1]{Corresponding author: Yuning Jiang}
		

		\address[a]{School of Vehicle and Mobility, Tsinghua University, 100084 Beijing, China}
		\address[b]{Automatic Control Laboratory, EPFL, 1024 Lausanne, Switzerland}
		
\begin{abstract}
Cooperative control of connected and automated vehicles (CAVs) promises smoother traffic flow. {In mixed traffic, where human-driven vehicles with unknown dynamics coexist, data-driven predictive control techniques allow for CAV safe and optimal control with measurable traffic data. However, the centralized control setting in most existing strategies limits their scalability for large-scale mixed traffic flow. To address this problem, this paper proposes a  cooperative \method{DeeP-LCC} (Data-EnablEd Predictive Leading Cruise Control) formulation and its distributed implementation algorithm. In cooperative \method{DeeP-LCC}, the traffic system is naturally partitioned into multiple subsystems with one single CAV, which collects local trajectory data for subsystem behavior predictions based on the Willems' fundamental lemma. Meanwhile, the cross-subsystem interaction is formulated as a coupling constraint. Then, we employ the Alternating Direction Method of Multipliers (ADMM) to design the distributed \method{DeeP-LCC} algorithm.} This algorithm achieves both computation and communication efficiency, as well as trajectory data privacy, through parallel calculation. Our simulations on different traffic scales verify the real-time wave-dampening potential of distributed \method{DeeP-LCC}, which can reduce fuel consumption by over $31.84\%$ in a large-scale traffic system of $100$ vehicles with only $5\%-20\%$ CAVs.

%
%

\end{abstract}
		
		
	\end{frontmatter}
	
\section{Introduction}

Traffic instabilities, in the form of periodic acceleration and deceleration of individual vehicles, cause a great loss of travel efficiency and fuel economy. This phenomenon, also known as traffic waves, is expected to be extensively eliminated with the emergence of connected and automated vehicles (CAVs). Particularly, with the advances of vehicle automation and wireless communications, CAV cooperative control promises system-wide traffic optimization and coordination, contributing to enhanced traffic mobility. One typical technology is Cooperative Adaptive Cruise Control (CACC), which organizes a 
group of CAVs into a single-lane platoon and maintains desired spacing and harmonized velocity, with dissipation of undesired traffic perturbations~\cite{li2017dynamical,zheng2016stability,milanes2013cooperative}. 

Despite the highly recognized potential of CAV cooperative control in both academy and industry, existing research mostly focuses on the fully-autonomous scenario with pure CAVs. For real-world implementation, however, the transition phase of mixed traffic with the coexistence of human-driven vehicles (HDVs) and CAVs may last for decades, making mixed traffic a more predominant pattern~\cite{stern2018dissipation,zheng2020smoothing,li2022cooperative}. By explicitly considering the behavior of surrounding HDVs that are under human control, recent research has revealed the potential of bringing significant traffic improvement with only a few CAVs. {Essentially, the CAVs can be utilized as mobile actuators for traffic control, leading to a recent notion of traffic Lagrangian control~\cite{stern2018dissipation,vinitsky2018lagrangian}. In a closed ring-road setup, the seminal real-world experiment in~\cite{stern2018dissipation}, followed by a series of theoretical analysis~\cite{xie2018heterogeneous,wang2020controllability} and simulation reproductions~\cite{wu2021flow,zheng2020analyzing}, reveals the capability of one single CAV in stabilizing the entire mixed traffic flow. }

{
Along this direction, multiple strategies have been designed for traffic-oriented CAV control in mixed traffic flow. One typical method is jam-absorption driving (JAD), which adjusts the CAV motion to leave enough inter-vehicle spacing for the traffic wave to be dissipated~\cite{he2016jam,nishi2013theory}. By extending the typical CACC frameworks to the mixed traffic setup, 
Connected Cruise Control (CCC) makes control decisions for one CAV at the tail by considering the motion of one or multiple HDVs ahead~\cite{orosz2016connected,jin2017optimal}. By enabling the CAV to respond to one HDV behind, the recent work in~\cite{molnar2020open} proposes a closed-loop traffic control paradigm to stabilize the upstream
traffic. Further, Leading Cruise Control (LCC)~\cite{wang2021leading} extends the idea in~\cite{orosz2016connected,jin2017optimal,molnar2020open} to a more general case by incorporating both the HDVs behind and ahead of the CAV into the system framework, and indicates that one CAV can not only adapt to the downstream traffic flow as \textit{a follower}, but also actively regulate the motion of the upstream traffic participants as \textit{a leader}.} The aforementioned work mostly focuses on the single-CAV case. When multiple CAVs coexist, the very recent work~\cite{li2022cooperative} reveals that rather than organizing all the CAVs into a platoon, one can allow CAVs to be naturally and arbitrarily distributed in mixed traffic and apply cooperative control decisions, contributing to greater traffic benefits.

\subsection{Data-Driven and Distributed Control for Mixed Traffic}

Essentially, mixed traffic is a complex human-in-the-loop cyber-physical system. For longitudinal control of CAVs in mixed traffic, one typical approach is to employ the well-known car-following model, \eg, the optimal velocity model (OVM)~\cite{bando1995dynamical} and the intelligent driver model (IDM)~\cite{treiber2000congested}, to describe the driving behavior of HDVs. Lumping the dynamics of CAVs and HDVs together, a parametric model of the entire mixed traffic system can be derived, allowing for model-based controller design. { Based on CCC/LCC-type frameworks, multiple model-based methods have been employed to enable CAVs to dissipate traffic waves, such as optimal control~\cite{wang2020controllability,jin2017optimal,wang2021optimal}, $\mathcal{H}_\infty$ control~\cite{zhou2020stabilizing,di2019cooperative,mousavi2021output} and model predictive control (MPC)~\cite{feng2021robust,gong2018cooperative,guo2021anticipative}. }{ These model-based methods require prior knowledge of mixed traffic dynamics for controller synthesis and parameter tuning. In practical traffic flow, however, it is non-trivial to accurately identify the driving behavior of one particular HDV, which tends to be uncertain and stochastic due to human nature. }

To address this problem, model-free approaches that circumvent the model identification process in favor of data-driven techniques have received increasing attention.  {Reinforcement learning~\cite{vinitsky2018lagrangian,wu2021flow,kreidieh2018dissipating} and adaptive dynamic programming~\cite{gao2016data,huang2020learning}, for example, have shown their potential in learning CAVs' wave-dampening strategies in mixed traffic flow.} Nevertheless, their lack of interpretability, sample efficiency and safety guarantees remains of primal concern~\cite{recht2019tour}. On the other hand, by integrating learning methods with MPC---a prime methodology for constrained optimal control problems, data-driven predictive control techniques provide a significant opportunity for reliable safe control with available data. {Following this idea, several methods have been applied for CAV control in mixed traffic, such as data-driven reachability analysis~\cite{lan2021data} and Koopman operator theory~\cite{zhan2022data}.} Very recently, Data-EnablEd Predictive Leading Cruise Control (\method{DeeP-LCC})~\cite{wang2022deeplcc}, which combines Data-EnablEd Predictive Control (DeePC)~\cite{coulson2019data} with LCC~\cite{wang2021leading}, directly utilizes measurable traffic data to design optimal CAV control inputs with collision-free considerations. { Both small-scale traffic simulations~\cite{wang2022deeplcc,wang2022data} and real-world miniature experiments~\cite{wang2022implementation} have validated its capability in mitigating traffic waves and improving fuel economy. }

{Despite the effectiveness of the aforementioned model-free methods, one common issue that has significantly prohibited their implementation is the centralized control setting. A central unit is deployed to gather all the available data, and assign control actions for each CAV. For large-scale mixed traffic systems with multiple CAVs and HDVs, this process is non-trivial to be completed during the system's sampling period given the potential delay in both wireless communications and online computations~\cite{negenborn2014distributed}. As discussed in~\cite{wang2022deeplcc}, the number of offline pre-collected data samples for centralized \method{DeeP-LCC} grows in a quadratic relationship when the traffic system scales up, leading to a dramatic increase in online computation burden. Moreover, due to the free joining or leaving maneuvers of individual vehicles (particularly those HDVs under human control), the flexible structure of the mixed traffic system, \ie, the spatial formation and penetration rates of CAVs~\cite{li2022cooperative}, could raise significant concerns about the excessive burden of recollecting traffic data and relearning CAV strategies. }

As an alternative, distributed control and optimization techniques are believed to be more scalable and feasible for large-scale traffic control. 
One particular method is the well-established Alternating Direction Method of Multipliers (ADMM)~\cite{boyd2011distributed}, which separates a large-scale optimization problem into smaller pieces that are easier to handle. Thanks to its efficient distributed optimization design with guaranteed convergence properties for convex problems, ADMM has seen wide applications in multiple areas, such as distributed learning~\cite{huang2019dp}, power control~\cite{erseghe2014distributed}, and wireless communications~\cite{zhou2019energy}. Given the multi-agent nature of traffic flow dynamics, which consists of the motion of multiple individual vehicles, ADMM has also been widely employed for CAV coordination in traffic flow, by solving local control problems and sharing information via vehicle-to-vehicle (V2V) or vehicle-to-everything (V2X) interactions; see, \eg,~\cite{li2020synchronous,zhang2021semi,li2021distributed}. 
{To our best knowledge, however, there has been limited research on data-driven distributed control for CAVs in the case of large-scale mixed traffic flow, with a very recent exception in~\cite{zhan2022data}, which combines Koopman operator theory with ADMM. Rather than offline training a neural network as in~\cite{zhan2022data}, this paper aims to develop ADMM-based data-driven distributed control algorithms through the well-established Willems' fundamental lemma~\cite{willems2005note}, which directly relies on measurable data for online behavior predictions.}

\subsection{Contributions}

\begin{figure*}[t]
	\centering
	\includegraphics[width=16.5cm]{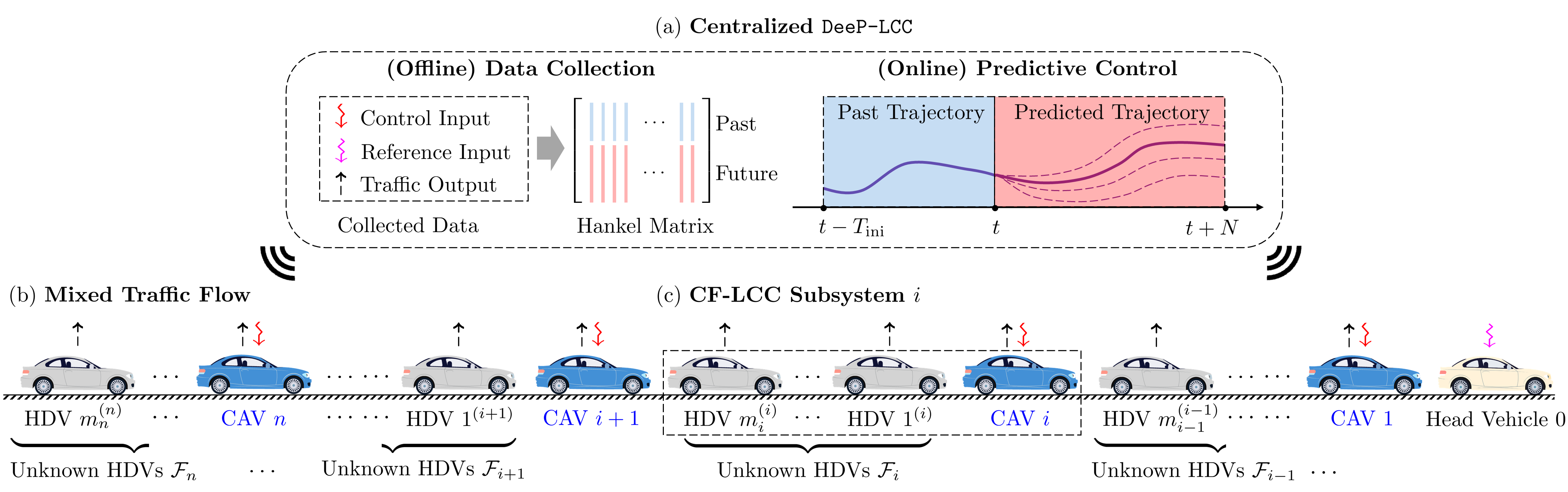}
	\caption{Schematic of centralized \method{DeeP-LCC} for CAVs in mixed traffic. (a) Centralized \method{DeeP-LCC}. \method{DeeP-LCC} collects the measurable data from the entire mixed traffic system, including traffic output, control input of the CAVs, and reference input, \ie, the velocity error of the head vehicle. Then, it utilizes these data to construct Hankel matrices for future trajectory predictions, and design the optimal future trajectory. More details can be found in~\cite{wang2022deeplcc}. (b) Mixed traffic scenario. The head vehicle is located at the beginning, indexed as 0, behind which there exist $n$ CAVs and $m$ HDVs. Between CAV $i$ and CAV $i+1$, there exist $m_i$ HDVs ($m_i \geq 0$). (c) {CF-LCC (Car-Following Leading Cruise Control)} subsystem $i$, consisting of a leading CAV $i$ and the following $m_i$ HDVs. More details can be found in~\cite[Section II-C]{wang2021leading}. 
	}
	\label{Fig:SystemSchematic}
\end{figure*}

Based on the centralized \method{DeeP-LCC} formulation~\cite{wang2022deeplcc}, this paper proposes a cooperative \method{DeeP-LCC} strategy for CAVs in large-scale mixed traffic flow, and presents its distributed implementation algorithm via ADMM. As illustrated in Fig.~\ref{Fig:SystemSchematic}(b), we consider an arbitrary setup of mixed traffic pattern, where there might exist multiple CAVs and HDVs with arbitrary spatial formations~\cite{li2022cooperative}. With local measurable data for each CAV and bidirectional topology~\cite{zheng2016stability} in CAV communications, our method allows CAVs to make cooperative control decisions to reduce traffic instabilities and mitigate traffic waves in a distributed manner. No prior knowledge of HDVs' driving dynamics are required, and safe and optimal guarantees are achieved. Precisely, the contributions of this work are as follows.

We first present a cooperative \method{DeeP-LCC} formulation with local data for large-scale mixed traffic control. Instead of establishing a data-centric representation for the entire mixed traffic system~\cite{wang2022deeplcc}, we naturally partition it into multiple {CF-LCC (Car-Following Leading Cruise Control)} subsystems~\cite{wang2021leading}, with one leading CAV and multiple HDVs following behind~(if they exist); see Fig.~\ref{Fig:SystemSchematic}(c) for illustration of one CF-LCC subsystem. Each CAV directly utilizes measurable traffic data from its own CF-LCC subsystem to design safe and optimal control behaviors. The interaction between neighbouring subsystems is formulated as a coupling constraint. In the case of linear dynamics  with noise-free data, it is proved that cooperative \method{DeeP-LCC} provides the identical optimal control performance compared with centralized \method{DeeP-LCC}~\cite{wang2022deeplcc}. For practical implementation, however, cooperative \method{DeeP-LCC} requires considerably fewer local data for each subsystem.

We then propose a tailored ADMM based distributed implementation algorithm (distributed \method{DeeP-LCC}) to solve the cooperative \method{DeeP-LCC} formulation. Particularly, we decompose the coupling constraint between neighbouring CF-LCC subsystems by introducing a new group of decision variables. In addition, via casting input/output constraints as trivial projection problems, the algorithm can be implemented quite efficiently, leaving no explicit optimization problems to be numerically solved. A bidirectional information flow topology---common in the pure-CAV platoon setting~\cite{zheng2016stability}---is needed for the ADMM iterations. Each CF-LCC subsystem exchanges only temporary computing data with its neighbours, contributing to V2V/V2X communication efficiency and local trajectory data privacy.

Finally, we carry out two different scales of traffic simulations to validate the performance of distributed \method{DeeP-LCC}. The moderate-scale experiment ($15$ vehicles with $5$ CAVs) shows that distributed \method{DeeP-LCC} could cost much less computing time than centralized \method{DeeP-LCC}, with a suboptimal performance in smoothing traffic flow. The experiment on the large-scale mixed traffic system ($100$ vehicles with $5\%-20\%$ CAVs), where the computation time for centralized \method{DeeP-LCC} is completely unacceptable, further verifies the capability and scalability of distributed \method{DeeP-LCC} in real-time mitigating traffic waves, saving over $31.84\%$ fuel consumption.

\subsection{Paper Organization and Notations}
The rest of this paper is organized as follows. Section~\ref{Sec:2} presents the input/output data of mixed traffic flow, and Section~\ref{Sec:3} briefly reviews the previous results on centralized \method{DeeP-LCC}. Section~\ref{Sec:4} presents the cooperative \method{DeeP-LCC} formulation, and Section~\ref{Sec:5} provides a tailored ADMM based distributed \method{DeeP-LCC} algorithm. Traffic simulations are discussed in Section~\ref{Sec:6}, and Section~\ref{Sec:7} concludes this paper.

\emph{Notation:}
We denote $\mathbb{N}$ as the set of all natural numbers, $\mathbb{N}_i^j$ as the set of natural numbers in the range of $[i,j]$ with $i \leq j$, $\mathbb{0}_n$ as a zero vector of size $n$, $\mathbb{0}_{m\times n}$ as~a~zero matrix of  size $m \times n$, and $I_n$ as an identity matrix of size $n \times n$. For a vector $a$ and a symmetric positive definite matrix $X$, $\|a\|_{X}^{2}$ denotes the quadratic form $a^\top X a$. Given vectors $a_1,a_2,\ldots,a_m$, we denote $\col(a_1,a_2,\ldots,a_m)=\begin{bmatrix}
a_1^\top,a_2^\top,\ldots,a_m^\top
\end{bmatrix}^\top$. Given matrices of the same column size $A_1,A_2,\ldots,A_m$, we denote $\col(A_1,A_2,\ldots,A_m)=\begin{bmatrix}
A_1^\top,A_2^\top,\ldots,A_m^\top
\end{bmatrix}^\top$. Denote $\diag(x_1,\ldots,x_m)$ as a diagonal matrix with $x_1,\ldots,x_m$ on its diagonal entries, and $\diag(D_1,\ldots,D_m)$ as a block-diagonal matrix with matrices $D_1,\ldots,D_m$ on its diagonal blocks. Finally, $\otimes$ represents the Kronecker product. 

\section{Input/Output Definition of Mixed Traffic Flow}
\label{Sec:2}

Consider a general single-lane mixed traffic system shown in Fig.~\ref{Fig:SystemSchematic}(b), where there exist one head vehicle, $n$ CAVs, and $m$ HDVs. The head vehicle, indexed as vehicle $0$, represents the vehicle immediately ahead of the first CAV. The CAVs are indexed as $1, 2, \ldots, n$ from front to end. Behind CAV~$i$ ($i\in\mathbb{N}_1^n$), there might exist $m_i$ ($m_i \geq 0$, $\sum_{i=1}^{n}{m_i} = m$) HDVs, and they are indexed as $1^{(i)},2^{(i)},\ldots,m_i^{(i)}$ in sequence. We introduce the following notations for the set consisting of vehicle indices---$\Omega$: all the vehicles; $\mathbb{N}_1^n$: all the CAVs; $\mathcal{F}$: all the HDVs; $\mathcal{F}_i$: those HDVs following behind CAV~$i$. Precisely, we have 
\begin{equation*}
    \mathcal{F}_i =\{1^{(i)},2^{(i)},\ldots,m_i^{(i)}\}, \;\; i \in  \mathbb{N}_1^n;\quad
    \mathcal{F} = \mathcal{F}_1 \cup \mathcal{F}_2 \cup \cdots \cup \mathcal{F}_n; 
    \quad \Omega = \mathbb{N}_1^n \cup \mathcal{F}.
\end{equation*}
Note that HDV~$m_i^{(i)}$ (if it exists) is the vehicle immediately ahead of CAV~$i+1$, $i\in \mathbb{N}_1^{n-1}$, and HDV~$m_n^{(n)}$ represents the last vehicle in this mixed traffic system. {It is assumed in this paper that all the vehicles, including CAVs and HDVs, are connected to the V2X communication network. Particularly, the velocity signal of the HDVs can be acquirable by the cloud unit in the centralized framework, or the CAVs in the distributed framework. Nevertheless, all the results can be generalized to the case where partial HDVs have connected capabilities, which will be discussed later.} %

\subsection{Definition for Input, Output and State}

We first specify the measurable output signals in the mixed traffic system. Denote the velocity and spacing of vehicle $i$ ($i \in \Omega$) at time $t$ as $v_i(t)$ and $s_i(t)$, respectively. To achieve wave mitigation in mixed traffic, the CAVs need to stabilize the traffic flow at a certain equilibrium state, where each vehicle moves with an identical equilibrium velocity $v^*$, \ie, $v_i(t)=v^*,\, i \in \Omega$, whilst maintaining an equilibrium spacing $s^*$. Here for simplicity, we use a homogeneous value $s^*$, but it could be varied for different vehicles.

Define the error states, including velocity errors $\tilde{v}_i(t)$ and spacing errors $\tilde{s}_i(t)$ from the equilibrium, as follows
\begin{equation} \label{Eq:ErrorState}
\tilde{v}_i(t)=v_i(t)-v^*,\; 
    \tilde{s}_i(t)=s_i(t)-s^*,\;\;i \in \Omega,
\end{equation}

{It is worth noting that in practice, not all the error states in~\eqref{Eq:ErrorState} are directly measurable. The raw data from V2X communications are mostly absolute position and velocity signals. To estimate the equilibrium velocity $v^*$ and obtain the velocity errors $\tilde{v}_i(t),\;i \in \Omega$, one approach is to utilize the average historical velocity of the head vehicle~\cite{wang2022implementation,jin2018connected}. For the equilibrium spacing $s^*$, this value for the HDVs is generally unknown and even time-varying; in contrast, the equilibrium spacing for the CAVs is designed by users, similarly to the desired spacing in typical CACC systems~\cite{li2017dynamical}. Accordingly, by appropriate design of the equilibrium spacing, the spacing errors for the CAVs $\tilde{s}_i(t),\; i\in\mathbb{N}_1^n$ can be obtained. In this paper, we assume that the velocity of all the vehicles can be acquired via V2X communication, and the spacing of all the CAVs can be obtained via on-board sensors. Then, the measurable signals are lumped into the aggregate output vector $y(t)$ of the mixed traffic system, given by
\begin{equation} \label{Eq:SystemOutput}
    y(t)=\col \left( y_1(t), y_2(t), \ldots, y_n(t) \right) \in \mathbb{R}^{2n+m},
\end{equation}
where
\begin{equation} \label{Eq:SubSystemOutput}
    y_i(t) = \begin{bmatrix}
    \tilde{v}_{i}(t), \tilde{v}_{1^{(i)}}(t),\ldots,\tilde{v}_{m_i^{(i)}}(t),\tilde{s}_{i}(t)
    \end{bmatrix}^{\top} \in \mathbb{R}^{m_i+2},\;\; i \in \mathbb{N}_1^n .
\end{equation}
This output $y(t)$ contains the velocity errors of all the vehicles and the spacing errors of only the CAVs. }{Regarding the spacing errors of the HDVs, some existing studies typically assume that these signals are also acquirable; see, \eg,~\cite{zheng2020smoothing,jin2017optimal,di2019cooperative,gao2016data,huang2020learning}.} 
Then, they consider the underlying state vector, defined as 
\begin{equation} \label{Eq:SystemState}
    x(t)=\col \left( x_1(t), x_2(t), \ldots, x_n(t) \right) \in \mathbb{R}^{2n+2m},
\end{equation}
where
\begin{equation} \label{Eq:SubSystemState}
    x_i(t)  =  \begin{bmatrix}
    \tilde{v}_{i}(t),  \tilde{v}_{1^{(i)}}(t), \ldots, \tilde{v}_{m_i^{(i)}}(t), \tilde{s}_{i}(t), \tilde{s}_{1^{(i)}}(t), \ldots, \tilde{s}_{m_i^{(i)}}(t)
    \end{bmatrix}^{\top}\in \mathbb{R}^{2+2m_i},\;\; i \in \mathbb{N}_1^n ,
\end{equation}
and design state-feedback control strategies.
This is impractical since the equilibrium spacing of the HDVs is indeed unknown in real traffic flow. In addition, note that the state~\eqref{Eq:SystemState} and the output~\eqref{Eq:SystemOutput} are transformed from those in~\cite{wang2022deeplcc} via row permutation, which has no influence on the fundamental system properties, such as controllability and observability. 

We next introduce the input signals in mixed traffic control. Denote the control input of each CAV as $u_i(t),\;i\in \mathbb{N}_1^n$, which could be the desired or actual acceleration of the CAVs~\cite{zheng2020smoothing,jin2017optimal,huang2020learning}. Lumping all the CAVs' control inputs, we define the aggregate control input of the entire mixed traffic system as 
\begin{equation} \label{Eq:ControlInput}
u(t) = \begin{bmatrix}u_{1}(t), u_{2}(t), \ldots,         u_{n}(t)\end{bmatrix}^{\top} \in \mathbb{R}^{n}.
\end{equation}
{Finally, the velocity error of the head vehicle from the equilibrium velocity $v^*$ is regarded as an external reference input signal $\epsilon (t)\in \mathbb{R}$ into the mixed traffic system,} given by 
\begin{equation} \label{Eq:ExternalInput}
    \epsilon (t)=\tilde{v}_{0}(t)=v_0(t)-v^* \in \mathbb{R}.
\end{equation}

\subsection{Parametric Model of Mixed Traffic Flow}
    
Based on the definitions of system state~\eqref{Eq:SystemState}, input~\eqref{Eq:ControlInput},~\eqref{Eq:ExternalInput} and output~\eqref{Eq:SystemOutput}, model-based strategies from the literature~\cite{jin2017optimal,zheng2020smoothing,wang2020controllability,zhou2020stabilizing} typically establish a parametric mixed traffic model for CAV controller design. They rely on a car-following model, \eg, IDM~\cite{kesting2010enhanced} or OVM~\cite{bando1995dynamical}, to describe the driving dynamics of HDVs, {whose general form can be written as
\begin{equation}
\label{Eq:HDVModel}
\dot{v}_i(t)=F_i\left(s_i(t),\dot{s}_i(t),v_i(t)\right) ,\;\;i \in \mathcal{F},
\end{equation}
where $\dot{s}_i(t)=v_{i-1}(t)-v_i(t)$ denotes the relative velocity of HDV $i$, and function $F_i(\cdot)$ represents the car-following dynamics of the HDV index as $i$. In general, the HDVs have heterogeneous behaviors, which indicates that $F_i(\cdot)$ could be different for different HDVs.} Assume that the CAVs' acceleration is utilized as the control input, \ie, \begin{equation} \label{Eq:CAVModel}
    \dot{v}_i(t)=u_i(t) ,\;\;i \in \mathbb{N}_1^n.
\end{equation}
Through linearization around equilibrium ($v^*,s^*$), a linearized state-space model for the mixed traffic system can be obtained via combining the driving dynamics~\eqref{Eq:HDVModel} and~\eqref{Eq:CAVModel} of each individual vehicle, which is in the form of~\cite{wang2022deeplcc,wang2022data}
\begin{equation} \label{Eq:DT_TrafficModel}
\begin{aligned}
x(k+1) =\;& Ax(k) + Bu(k) + H \epsilon(k),\\
y(k) =\;& Cx(k),
\end{aligned}
\end{equation}
where $A,B,C,H$ are system matrices of compatible dimensions; see~\cite[Section II]{wang2022deeplcc} for a specific representation under the LCC framework with a homogeneous assumption for the HDV behaviors, \ie, the function $F_i(\cdot)$ in~\eqref{Eq:HDVModel} is the same for all the HDVs. 

In real traffic flow, the human driving dynamics~\eqref{Eq:HDVModel} for individual vehicles are non-trivial to identify. Thus, the mixed traffic model~\eqref{Eq:DT_TrafficModel} is practically unknown. To address this problem, the recently proposed \method{DeeP-LCC} method circumvents the necessity of identifying an HDV's car-following dynamics; instead, it employs a data-centric non-parametric representation from input/output traffic data for behavior prediction and controller design. This method directly relies on the collected data, and the following definition of persistent excitation~\cite{willems2005note} is needed.
\begin{definition}
	Let $T,\,l\in \mathbb{N}$ and $l \leq T$. Define the Hankel matrix of order $l$ for the signal sequence  $\omega = \col (\omega(1),\omega(2),$ $\ldots,\omega(T))$ as
	\begin{equation}
		\mathcal{H}_{l}(\omega):=\begin{bmatrix}
			\omega(1) &\omega(2) & \cdots & \omega(T-l+1) \\
			\omega(2) &\omega(3) & \cdots & \omega(T-l+2) \\
			\vdots & \vdots & \ddots & \vdots \\
			\omega(l) &\omega(l+1) & \cdots & \omega(T)
		\end{bmatrix}.
	\end{equation}
	We call $\omega$ persistently exciting of order $l$ if the Hankel matrix $\mathcal{H}_{l}(\omega)$ has full row rank.   
\end{definition}

\section{Review of Centralized \method{DeeP-LCC} Formulation}
\label{Sec:3}

In this section, we briefly introduce the \method{DeeP-LCC} strategy from~\cite{wang2022deeplcc,wang2022data}, which is formulated in a centralized control setting, as illustrated in Fig.~\ref{Fig:SystemSchematic}(a). Essentially, \method{DeeP-LCC} is adapted from the standard DeePC method~\cite{coulson2019data}, which has seen wide applications in multiple fields, such as quadcopter systems~\cite{elokda2021data}, power grids~\cite{huang2021decentralized}, and building control~\cite{lian2023adaptive}.

\subsection{Data-Centric Representation of Mixed Traffic Flow}

For the data-centric representation of the mixed traffic system~\eqref{Eq:DT_TrafficModel}, we begin by collecting an input/output data sequence of length $T$ from this system:

\vspace{0.2em}
\noindent\textbf{Centralized Data:}
\begin{subequations} \label{Eq:CollectedData}
\begin{align}
u^\mathrm{d}&=\col \left(u^\mathrm{d}(1),u^\mathrm{d}(2),\ldots,u^\mathrm{d}(T)\right)\in \mathbb{R}^{nT},\\
\epsilon^\mathrm{d}&=\col \left(\epsilon^\mathrm{d}(1),\epsilon^\mathrm{d}(2),\ldots,\epsilon^\mathrm{d}(T)\right) \in \mathbb{R}^{T},\\
y ^\mathrm{d}&=\col \left(y^\mathrm{d}(1),y^\mathrm{d}(2),\ldots,y^\mathrm{d}(T)\right) \in \mathbb{R}^{(2n+m)T}.
\end{align}
\end{subequations}
Then, let $T_{\mathrm{ini}},N \in \mathbb{N}$ ($T_{\mathrm{ini}}+N \leq T$), and define 
\begin{equation}
\label{Eq:DataHankel}
\begin{gathered}
\begin{bmatrix}
U_{\mathrm{p}} \\
U_{\mathrm{f}}
\end{bmatrix}:=\mathcal{H}_{T_{\mathrm{ini}}+N}(u^{\mathrm{d}}), \quad \begin{bmatrix}
E_{\mathrm{p}} \\
E_{\mathrm{f}}
\end{bmatrix}:=\mathcal{H}_{T_{\mathrm{ini}}+N}(\epsilon^{\mathrm{d}}), \quad  
\begin{bmatrix}
Y_{\mathrm{p}} \\
Y_{\mathrm{f}}
\end{bmatrix}:=\mathcal{H}_{T_{\mathrm{ini}}+N}(y^{\mathrm{d}}),
\end{gathered}
\end{equation}
where $U_{\mathrm{p}}$ and $U_{\mathrm{f}}$ consist of the first $T_{\mathrm{ini}}$ block rows and the last $N$ block rows of $\mathcal{H}_{T_{\mathrm{ini}}+N}(u^{\mathrm{d}})$, respectively (similarly for $E_{\mathrm{p}}, E_{\mathrm{f}}$ and $Y_{\mathrm{p}}, Y_{\mathrm{f}}$). The partition in~\eqref{Eq:DataHankel} separates each column in the data Hankel matrix, which is a consecutive data sequence of length $T_\mathrm{ini}+N$, into $T_\mathrm{ini}$-length past data and $N$-length future data.

\begin{assumption}[Persistent excitation in a centralized setup] \label{Assumption:PersistentExcitation}
The pre-collected control input data sequence $u^\mathrm{d}$ is persistently exciting of order $T_\mathrm{ini}+N+2n+2m$.
  
\end{assumption}

Given the persistent excitation of pre-collected data and the controllability and observability assumption for the underlying system (see~\cite[Theorem 1]{wang2022deeplcc} for a mild condition in the linearized setup), we have the following data-centric representation of mixed traffic behavior.


\begin{lemma}[{\cite[Proposition 2]{wang2022deeplcc}}]
\label{Proposition:DeePCMixedTraffic}
Let $t\in \mathbb{N}$ be the current time, and denote the most recent $T_\mathrm{ini}$-length past control input sequence $u_\mathrm{ini}$ and the future $N$-length control input sequence $u$ as
\begin{subequations} \label{Eq:TrajectoryDefinition}
\begin{align}
u_{\mathrm{ini}}&=\col\left(u(t-T_{\mathrm{ini}}),u(t-T_{\mathrm{ini}}+1),\ldots,u(t-1)\right),\\
u&= \col\left(u(t),u(t+1),\ldots,u(t+N-1)\right),
\end{align}
\end{subequations} 
respectively (similarly for $\epsilon_\mathrm{ini},\epsilon$ and $y_\mathrm{ini},y$). By Assumption~\ref{Assumption:PersistentExcitation} and a controllability and observability assumption for the underlying system, the sequence $\col (u_\mathrm{ini},\epsilon_{\mathrm{ini}},y_\mathrm{ini},u,\epsilon,y)$ is a $(T_{\mathrm{ini}}+N)$-length trajectory of the linearized mixed traffic system~\eqref{Eq:DT_TrafficModel}, if and only if there exists a vector $g\in \mathbb{R}^{T-T_\mathrm{ini}-N+1}$, such that
\begin{equation}
\label{Eq:DeePLCCAchievability}
\begin{bmatrix}
U_\mathrm{p} \\ E_{\mathrm{p}}\\Y_\mathrm{p} \\ U_\mathrm{f} \\ E_{\mathrm{f}}\\ Y_\mathrm{f}
\end{bmatrix}g=
\begin{bmatrix}
u_\mathrm{ini} \\ \epsilon_{\mathrm{ini}}\\ y_\mathrm{ini} \\ u \\\epsilon \\ y
\end{bmatrix}.
\end{equation}
If $T_{\mathrm{ini}} \geq 2n+2m$, $y$ is uniquely determined from~\eqref{Eq:DeePLCCAchievability}, $\forall (u_\mathrm{ini} ,\epsilon_\mathrm{ini}, y_\mathrm{ini},u,\epsilon)$.   
\end{lemma}

{This lemma is adapted from the well-established Willems' fundamental lemma~\cite{willems2005note} and standard DeePC~\cite{coulson2019data}, which reveals that any valid trajectory of a controllable linear time-invariant (LTI) system can be constructed by a finite number of input/output data samples, provided sufficiently rich inputs during data collection. \method{DeeP-LCC} applies this result to mixed traffic control and replaces the need for a parametric mixed traffic model~\eqref{Eq:DT_TrafficModel} or identification process for human's driving behaviors. Furthermore, this result allows for future behavior prediction for $y$ under an assumed future input $u$ and reference input $\epsilon$, given pre-collected traffic data $(u^\mathrm{d},\epsilon^\mathrm{d},y^\mathrm{d})$ and the most recent past trajectory $(u_\mathrm{ini},\epsilon_\mathrm{ini},y_\mathrm{ini})$. Similar adaptations can also be found in recent work on power systems~\cite{huang2021decentralized} and building control~\cite{lian2023adaptive}.}



\subsection{Centralized Formulation of \method{DeeP-LCC}}

We utilize the non-parametric representation~\eqref{Eq:DeePLCCAchievability} for predictive control of CAVs in mitigating traffic waves in mixed traffic. Define 
\begin{equation} \label{Eq:CostDefinition}
V(y,u) = \sum\limits_{k=t}^{t+N-1}\left( \left\|y(k)\right\|_{Q}^{2}+\left\|u(k)\right\|_{R}^{2}\right),     
\end{equation}
as the cost function for the future trajectory from time $t$ to time $t+N-1$, which penalizes the traffic output and CAVs' control inputs with weight coefficients $Q$ and $R$, respectively. Precisely, define $w_v,w_s,w_u > 0$ as the weight coefficients for velocity errors, spacing errors and control inputs, respectively, and we have
\begin{equation}
    Q = \diag(Q_1,Q_2,\ldots,Q_n),
\end{equation}
where 
\begin{equation}
    Q_i = \diag(\overbrace{\omega_v,\omega_v,\ldots,\omega_v}^{m_i},\omega_s),\;\; i \in \mathbb{N}_1^n,
\end{equation}
and
\begin{equation}
    R = \diag(\overbrace{\omega_u,\omega_u,\ldots,\omega_u}^{n}).
\end{equation}
Then, we formulate the following optimization problem for predictive control of the linearized mixed traffic system~\eqref{Eq:DT_TrafficModel} with noise-free data~\eqref{Eq:CollectedData}: 

\vspace{0.2em}
\noindent\textbf{Centralized Linear \method{DeeP-LCC}:} 
\begin{subequations} \label{Eq:CentralizedDeePLCCLinearized}
\begin{align}
\min_{g,u,y} \quad &V(u,y) \\
\st \quad &\eqref{Eq:DeePLCCAchievability},\\
&\epsilon = \hat{\epsilon}, \label{Eq:CentralizedLinearEstimation}\\
& u\in \mathcal{U},y\in \mathcal{Y},
\end{align}
\end{subequations}
where $u\in \mathcal{U},y\in \mathcal{Y}$ represent the input/output constraints, and $\hat{\epsilon} \in\mathbb{R}^{N}$ denotes the estimation of the future reference input sequence $\epsilon$.  

In practical traffic flow, the consistency of the non-parametric behavior representation~\eqref{Eq:DeePLCCAchievability} is usually compromised  due to data noise and HDVs' nonlinear and non-deterministic behavior. Thus, the original optimization problem~\eqref{Eq:CentralizedDeePLCCLinearized} might have no feasible solutions. The following regularized version is proposed to obtain optimal control for CAVs in practical traffic flow~\cite{wang2022deeplcc}:

\vspace{0.2em}
\noindent\textbf{Centralized \method{DeeP-LCC}:} 
  \begin{subequations} \label{Eq:CentralizedDeePLCC}
 \begin{align}
 \min_{g,u,y,\sigma_y} \quad & J (y,u,g,\sigma_y)\\
 \st \quad & \begin{bmatrix}
 U_\mathrm{p} \\ E_{\mathrm{p}}\\Y_\mathrm{p} \\ U_\mathrm{f} \\ E_{\mathrm{f}}\\ Y_\mathrm{f}
 \end{bmatrix}g=
 \begin{bmatrix}
 u_\mathrm{ini} \\ \epsilon_{\mathrm{ini}}\\ y_\mathrm{ini} \\ u \\\epsilon \\ y
 \end{bmatrix}+\begin{bmatrix}
 0\\0\\ \sigma_y \\0 \\0 \\0
 \end{bmatrix},\\ &\epsilon = \hat{\epsilon}, \\
 & u\in \mathcal{U},y\in \mathcal{Y},
 \end{align}
 \end{subequations}
with
\begin{equation} \label{Eq:CentralizedCostFunction}
    J (y,u,g,\sigma_y) = V(u,y)+\lambda_g \left\|g\right\|_2^2+\lambda_y \left\|\sigma_y\right\|_2^2,
\end{equation}
where $\sigma_y \in \mathbb{R}^{(n+m)T_\mathrm{ini}}$ is a slack variable to ensure feasibility, and a sufficiently large weight coefficient $\lambda_y>0$ is introduced for penalization in the cost function, which allows for $\sigma_y \neq 0$ only if the equality constraint~\eqref{Eq:DeePLCCAchievability} is infeasible. A two-norm penalty on $g$ with $\lambda_g>0$ is also included in the cost function to avoid over-fitting of noisy data, and this regulation has been shown to coincide with distributional two-norm robustness~\cite{huang2021decentralized,coulson2019regularized}. 

\begin{remark}
\method{DeeP-LCC} requires a centralized cloud unit to collect data~\eqref{Eq:CollectedData} and $u_\mathrm{ini},\epsilon_\mathrm{ini},y_\mathrm{ini}$ of the entire mixed traffic system and assign control inputs for all the CAVs via solving the centralized optimization problem~\eqref{Eq:CentralizedDeePLCC}. Both traffic simulations~\cite{wang2022deeplcc} and real-world tests~\cite{wang2022implementation} have validated its potential in mitigating traffic waves in a moderate-scale setup. In larger-scale traffic flow, however, the computation time in solving~\eqref{Eq:CentralizedDeePLCC} would soon become intolerable for real-time implementation, and the communication constraints and flexible structures of mixed traffic patterns would also limit the practical application. 
\end{remark}

\section{Cooperative \method{DeeP-LCC}}
\label{Sec:4}

In this section, we first introduce the partition design of the entire mixed traffic system into CF-LCC subsystems, and present local  \method{DeeP-LCC} for each CF-LCC subsystem. Then, we show the cooperative \method{DeeP-LCC} formulation by coupling each CF-LCC subsystems, and present the theoretical analysis of the relationship between cooperative \method{DeeP-LCC} and centralized \method{DeeP-LCC}.

\subsection{CF-LCC Subsystem Partition and Local \method{DeeP-LCC}}

\begin{figure*}[t]
	\centering
	\includegraphics[width=16.5cm]{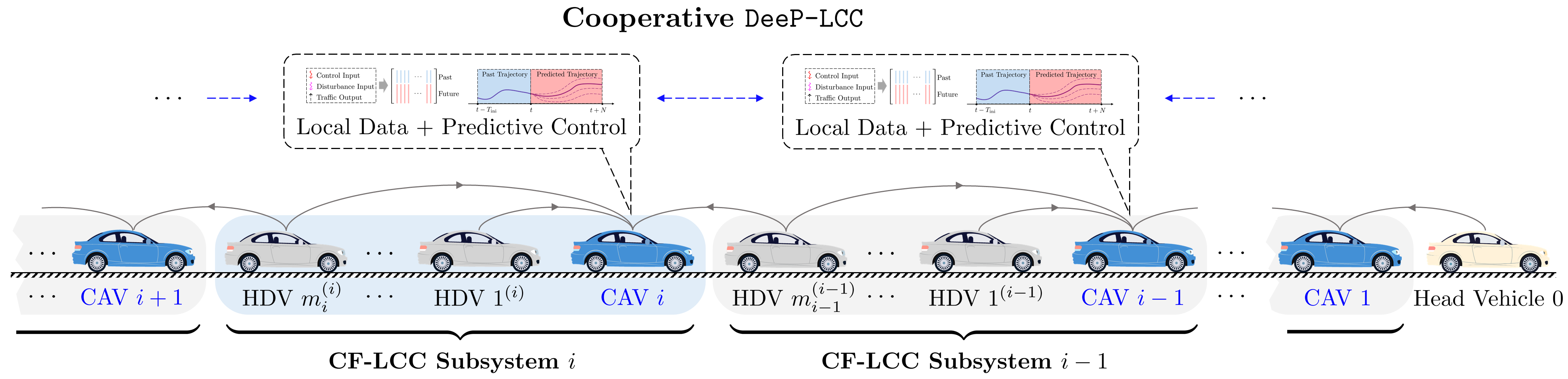}
	\caption{Schematic of partition of the entire mixed traffic system and cooperative \method{DeeP-LCC}. Each CF-LCC system consists of CAV $i$ and its following HDVs, where the CAV monitors the motion of those following HDVs and also the HDV immediately ahead, \ie, HDV $m^{(i-1)}_{i-1}$; see the gray solid arrows. In cooperative \method{DeeP-LCC}, only local data are needed to design the future behavior of each subsystem. Note that a bidirectional information flow typology (represented by blue dashed arrows) is designed for data exchange between neighbouring CF-LCC systems, which will be detailed in Section~\ref{Sec:4}.  }
	\label{Fig:CFLCCSubsystem}
\end{figure*}

Recall the general mixed traffic system with $n$ CAVs and $m$ HDVs as shown in Fig.~\ref{Fig:SystemSchematic}(b). The centralized \method{DeeP-LCC} formulation utilizes input/output data~\eqref{Eq:CollectedData} of the entire mixed traffic system to describe the traffic behavior, neglecting its inherent dynamics structure. Indeed, the mixed traffic system can be naturally partitioned into $n$ subsystems, consisting of CAV $i$ and its following $m_i$ HDVs represented by $\mathcal{F}_i=\{1^{(i)},2^{(i)},\ldots,m_i^{(i)}\}, \, i \in  \mathbb{N}_1^n$; see Fig.~\ref{Fig:CFLCCSubsystem} for demonstration. This subsystem is named as CF-LCC (Car-Following LCC) in~\cite{wang2021leading}, where one single CAV is leading the motion of the HDVs behind, whilst following a vehicle immediately ahead, which is HDV $m_{i-1}^{(i-1)}$ (or CAV $i-1$ if $\mathcal{F}_{i-1}=\varnothing$). If $m_i=0$ for some $i$, \ie, CAV $i$ has no following HDVs but CAV $i+1$ instead, this CAV itself stands as an independent CF-LCC subsystem. { 
In addition to our partition strategy, there exist alternative approaches for managing mixed traffic flow at scale. For instance, one can also group the HDVs ahead (indexed from $1^{(i-1)}$ to $m_{i-1}^{(i-1)}$) and the following CAV (indexed as $i$) into a subsystem; see, \eg,~\cite{di2019cooperative,zhan2022data}. It has been shown in~\cite{wang2021leading} that in a linearized subsystem containing one CAV, the states of the HDVs behind are controllable with respect to the CAV's input, making them explicit targets for optimization. However, the HDVs ahead of CAV $i$ are not under its direct influence (but CAV $i-1$ instead) and primarily serve to inform predictions about ahead traffic conditions.}

In the following we focus on each CF-LCC subsystem~$i$, which, compared to the general one in Fig.~\ref{Fig:SystemSchematic}(b), is essentially a smaller-size mixed traffic system with one head vehicle (HDV $m_{i-1}^{(i-1)}$), one CAV $i$, and $m_i$ HDVs. Therefore, the previous \method{DeeP-LCC} formulation can be directly applied to this subsystem. Precisely, for CF-LCC subsystem $i$ at time $t$, define system output as $y_i(t)$ in~\eqref{Eq:SubSystemOutput}, which contains the velocity errors of CAV $i$ and the following HDVs in $\mathcal{F}_i$, and denote the control input as $u_i(t)$, consistent with the corresponding entry in~\eqref{Eq:ControlInput}.  Additionally, we introduce 
\begin{equation} \label{Eq:ExternalInputSubsystem}
    \epsilon_i(t) = \tilde{v}_{m_{i-1}^{(i-1)}}(t)=  v_{m_{i-1}^{(i-1)}}(t)-v^* \in \mathbb{R},
\end{equation} 
as the reference input, which is the velocity error of the vehicle immediately ahead of CAV $i$. Note that $\epsilon_1$ is consistent with $\epsilon$ in~\eqref{Eq:ExternalInput}, representing the velocity error of the head vehicle at the very beginning of the entire mixed traffic system, \ie, vehicle $0$.

Similarly to~\eqref{Eq:DT_TrafficModel}, through linearization around equilibrium, a state-space model can be derived for each CF-LCC subsystem, in the form of 
\begin{equation} \label{Eq:DT_CFLCCModel}
\begin{cases}
x_i(k+1) = A_ix_i(k) + B_iu_i(k) + H_i \epsilon_i(k),\\
y_i(k) = C_ix_i(k),
\end{cases}
\end{equation}
where $A_i,B_i,H_i,C_i$ are system matrices of compatible dimensions; see~\cite[Section II-C]{wang2021leading} for a specific expression.

In practice, however, the system model might be unknown. Following the process in centralized \method{DeeP-LCC}, we can also obtain a data-centric representation for each CF-LCC subsystem. Precisely, collect an input/output data sequence of length $T_i\in \mathbb{N}$ for CF-LCC subsystem $i$:

\vspace{0.2em}
\noindent\textbf{Local Data:}
\begin{subequations} \label{Eq:CollectedDataSubsystem}
\begin{align}
u_i^\mathrm{d}&=\col \left(u_i^\mathrm{d}(1),u_i^\mathrm{d}(2),\ldots,u_i^\mathrm{d}(T_i)\right)\in \mathbb{R}^{T_i},\\
\epsilon_i^\mathrm{d}&=\col \left(\epsilon_i^\mathrm{d}(1),\epsilon_i^\mathrm{d}(2),\ldots,\epsilon_i^\mathrm{d}(T_i)\right) \in \mathbb{R}^{T_i},\\
y_i^\mathrm{d}&=\col \left(y_i^\mathrm{d}(1),y_i^\mathrm{d}(2),\ldots,y_i^\mathrm{d}(T_i)\right) \in \mathbb{R}^{(m_i+2)T_i}.
\end{align}
\end{subequations}
These data sequences are utilized to construct data Hankel matrices $U_{i,\mathrm{p}},U_{i,\mathrm{f}},E_{i,\mathrm{p}},E_{i,\mathrm{f}},Y_{i,\mathrm{p}},Y_{i,\mathrm{f}}$ by a similar procedure in~\eqref{Eq:DataHankel}, with a same time horizon $T_\mathrm{ini}$ and $N$.

With these pre-collected data for each subsystem, we have the following result motivated by Lemma~\ref{Proposition:DeePCMixedTraffic}. 

{
\begin{assumption}[Persistent excitation in a distributed setup] \label{Assumption:PersistentExcitationSubsystem}
The pre-collected control input data sequence $u_i^\mathrm{d}$ for each CF-LCC subsystem $i,\;i \in \mathbb{N}_1^n$ is persistently exciting of order $T_\mathrm{ini}+N+2+2m_i$. 
\end{assumption}
}

\begin{corollary}
Following the definition in~\eqref{Eq:TrajectoryDefinition}, for CF-LCC subsystem $i$ at the current time $t$, denote the most recent $T_\mathrm{ini}$-length past trajectory and the future $N$-length trajectory as $u_{i,\mathrm{ini}},\epsilon_{i,\mathrm{ini}},y_{i,\mathrm{ini}}$ and $u_i,\epsilon_i,y_i$, respectively.
By Assumption~\ref{Assumption:PersistentExcitationSubsystem} and a controllability and observability assumption for the underlying system, the sequence $\col (u_{i,\mathrm{ini}} ,\epsilon_{i,\mathrm{ini}},$ $ y_{i,\mathrm{ini}},u_i,\epsilon_i,y_i)$ is a $(T_{\mathrm{ini}}+N)$-length trajectory of the linearized CF-LCC subsystem~\eqref{Eq:DT_CFLCCModel}, if and only if there exists a vector $g_i\in \mathbb{R}^{T_i-T_\mathrm{ini}-N+1}$, such that
\begin{equation}
\label{Eq:DeePLCCAchievabilitySubsystem}
\begin{bmatrix}
 U_{i,\mathrm{p}} \\ E_{i,\mathrm{p}}\\Y_{i,\mathrm{p}} \\ U_{i,\mathrm{f}} \\ E_{i,\mathrm{f}}\\ Y_{i,\mathrm{f}}
 \end{bmatrix}g_i=
 \begin{bmatrix}
 u_{i,\mathrm{ini}} \\ \epsilon_{i,\mathrm{ini}}\\ y_{i,\mathrm{ini}} \\ u_i \\\epsilon_i \\ y_i
 \end{bmatrix}.
\end{equation}
If $T_{\mathrm{ini}} \geq 2+2m_i$, $y_i$ is uniquely determined from~\eqref{Eq:DeePLCCAchievabilitySubsystem}, $\forall (u_{i,\mathrm{ini}} ,\epsilon_{i,\mathrm{ini}}, y_{i,\mathrm{ini}},u_i,\epsilon_i)$.   
\end{corollary}

Based on the data-centric representation of CF-LCC subsystem, we can naturally present the local formulation of \method{DeeP-LCC}, motivated by~\eqref{Eq:CentralizedDeePLCCLinearized}.

\vspace{0.2em}
\noindent\textbf{Local Linear \method{DeeP-LCC}:}
\begin{subequations} \label{Eq:LocalDeePLCC}
 \begin{align}
 \min_{u_i,y_i} \quad & V_i(u_i,y_i)\\
 \st \quad & \eqref{Eq:DeePLCCAchievabilitySubsystem},\\ &\epsilon_i = \hat{\epsilon}_i,\\
 & u_i\in \mathcal{U}_i,y_i\in \mathcal{Y}_i,
 \end{align}
 \end{subequations}
 where
 \begin{equation} \label{Eq:CostDefinitionSubsystem}
V_i(u_i,y_i) = \sum\limits_{k=t}^{t+N-1}\left( \left\|y_i(k)\right\|_{Q_i}^{2}+\left\|u_i(k)\right\|_{\omega_u}^{2}\right).     
\end{equation}
In~\eqref{Eq:LocalDeePLCC}, $\hat{\epsilon}_i$ denotes the estimation of the future reference input, and $u_i\in \mathcal{U}_i,y_i\in \mathcal{Y}_i$ represent the input/output constraints for subsystem $i$. It is assumed that the input/output constraints for the CF-LCC subsystems are consistent with those for the entire mixed traffic system, \ie, we have 
\begin{equation} \label{Eq:ConstraintConsistency}
    u \in \mathcal{U},y\in \mathcal{Y} \iff u_i\in \mathcal{U}_i,y_i\in \mathcal{Y}_i,\,\forall i \in \mathbb{N}_1^n.
\end{equation}

\subsection{Cooperative \method{DeeP-LCC} via Coupling Constraints}

In problem~\eqref{Eq:LocalDeePLCC}, each CAV relies only on the local data for controller design. Inside the CF-LCC subsystem, each CAV needs to monitor the motion of the following HDVs and the motion of the vehicle immediately ahead; see the gray arrows in Fig.~\ref{Fig:CFLCCSubsystem}. Particularly, the control objective is limited to improving the performance of the local CF-LCC subsystem. Despite the possible capability in mitigating traffic waves by applying~\eqref{Eq:LocalDeePLCC} to each CAV in mixed traffic, a noncooperative behavior is expected, and thus the resulting performance might be compromised. Moreover, an accurate estimation of $\hat{\epsilon}_i$ is always non-trivial.

\begin{figure*}[t]
	\centering
	\includegraphics[width=11cm]{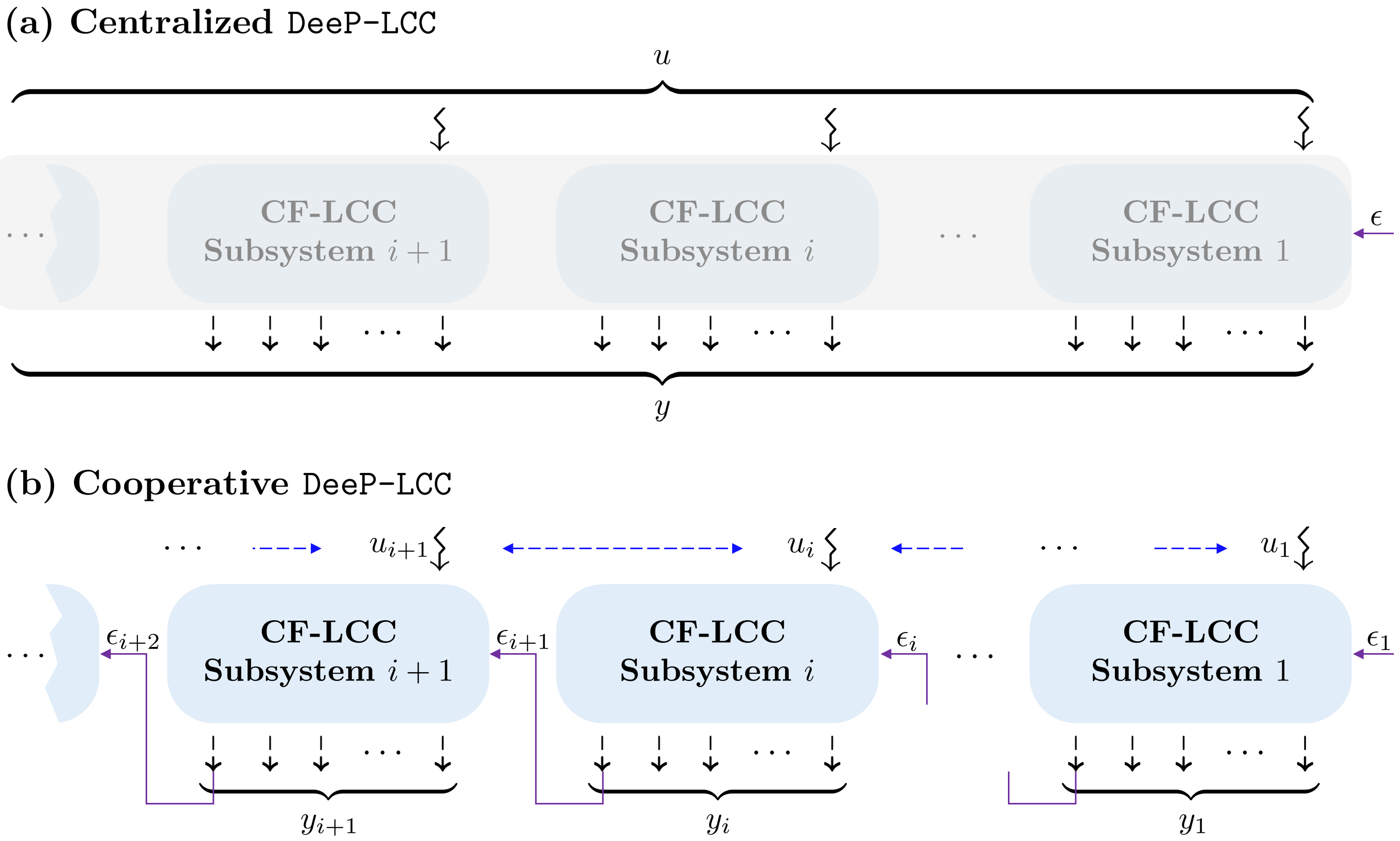}
	\caption{Schematic for comparison between centralized \method{DeeP-LCC} and cooperative \method{DeeP-LCC}. The magenta arrow denotes the fact that one output signal $\epsilon_{i+1}$ from the subsystem $i$ ahead acts as the reference input into the subsystem $i+1$ behind. The blue dashed arrows represent the bidirectional  information flow topology between neighbouring subsystems, which will be detailed in Section~\ref{Sec:4}.}
	\label{Fig:Comparison}
\end{figure*}

Necessary information exchange is needed to coordinate the control tasks between each CF-LCC subsystem in the mixed traffic flow. Indeed, as shown in~\eqref{Eq:ExternalInputSubsystem}, one of the output signals of subsystem $i$ (the velocity error of the last HDV $m_i^{(i)}$) acts as the reference input of subsystem $i+1$ (the velocity error of the vehicle ahead of CAV $i+1$). Therefore, each CF-LCC subsystem $i+1$ could receive the predicted output signal $y_i$ from CF-LCC subsystem $i$, and utilizes this signal to construct the estimation of future external reference $\hat{\epsilon}_{i+1}$; see Fig.~\ref{Fig:Comparison}(b) for illustration. Precisely, we have
\begin{equation} \label{Eq:CouplingDynamics}
    \hat{\epsilon}_{i+1} = \tilde{v}_{m_{i}^{(i)}} = K_i y_i  ,\;\; i\in \mathbb{N}_1^{n-1},
\end{equation}
where
$$
K_i = I_{N} \otimes \begin{bmatrix}
\mathbb{0}_{1 \times m_i} & 1 & 0
\end{bmatrix}.
$$
This information exchange between CF-LCC subsystems $i$ and $i+1$ facilitates a coordination behavior between neighbouring subsystems. Specifically, we can sum up the local optimization problem in~\eqref{Eq:LocalDeePLCC} and introduce~\eqref{Eq:CouplingDynamics} as a coupling constraint. Then, the following cooperative \method{DeeP-LCC} formulation can be established for the entire linearized mixed traffic system with noise-free data.

\vspace{0.2em}
\noindent\textbf{Cooperative Linear \method{DeeP-LCC}:} 
\begin{subequations} \label{Eq:CooperativeDeePLCCLinearized}
 \begin{align}
 \min_{ \substack{g_i,u_i,y_i  \\
      (i\in \mathbb{N}_1^{n})} } \quad & \sum_{i=1}^n V_i(u_i,y_i)\\
 \st \quad & \eqref{Eq:DeePLCCAchievabilitySubsystem} ,\;\;  i \in \mathbb{N}_1^{n},\\ 
 &\epsilon_{1} = \hat{\epsilon}, \label{Eq:CooperativeLinearEstimation}\\
 & \epsilon_{i+1} = K_i y_i ,\;\;i \in \mathbb{N}_1^{n-1}, \label{Eq:CooperativeLinearCoupling}
 \\ & u_i\in \mathcal{U}_i,y_i\in \mathcal{Y}_i ,\;\;  i \in \mathbb{N}_1^{n},
 \end{align}
 \end{subequations}
where the cost function is a summation of the local cost of each subsystem in the local formulation~\eqref{Eq:LocalDeePLCC}. Solving the cooperative \method{DeeP-LCC} formulation~\eqref{Eq:CooperativeDeePLCCLinearized} provides a cooperative behavior for all the subsystems. Comparing the cooperative formulation~\eqref{Eq:CooperativeDeePLCCLinearized} with the local formulation~\eqref{Eq:LocalDeePLCC}, there only exists one estimation for the reference input of the head vehicle (vehicle $0$) for subsystem $1$, as shown in~
\eqref{Eq:CooperativeLinearEstimation}, which is consistent with~\eqref{Eq:CentralizedLinearEstimation} in the centralized formulation~\eqref{Eq:CentralizedDeePLCCLinearized}.  In addition, the information exchange~\eqref{Eq:CooperativeLinearCoupling} acts as a coupling constraint in the optimization problem~\eqref{Eq:CooperativeDeePLCCLinearized}. 

\begin{remark}
Fig.~\ref{Fig:Comparison} illustrates the comparison between centralized \method{DeeP-LCC} and cooperative \method{DeeP-LCC}. In centralized \method{DeeP-LCC}, all the inputs and outputs are stacked together for behavior representation. In cooperative \method{DeeP-LCC}, by contrast, we partition the original large-scale system into multiple CF-LCC subsystems, and the local input and output data are collected for local system representation. Indeed, we assume the prior knowledge of system structure: one output signal from each subsystem $i$ acts as the reference input $\epsilon_{i+1} $ into the subsystem $i+1$ behind, as described in~\eqref{Eq:CouplingDynamics}.
\end{remark}

{
\begin{remark}
    In this paper, we assume that all the HDVs are connected to the V2X network. It is worth noting that \method{DeeP-LCC} directly relies on the input/output trajectory data for controller design, and when partial HDVs are connected, the measurable output signal will be changed. Indeed, it can be intuitively known from Fig.~\ref{Fig:Comparison} that the lowest connectivity requirement is that the HDV immediately ahead of each CAV, which is indexed as $m_i^{(i)}$, $i\in\mathbb{N}_1^n$ (\ie, the HDV at the tail of each CF-LCC subsystem $i$), must be connected. In this case, the output definition in~\eqref{Eq:SubSystemOutput} is degraded to $y_i(t) = \left[\tilde{v}_{i}(t), \tilde{v}_{m_i^{(i)}}(t),\tilde{s}_{i}(t)
    \right]^{\top}  \in \mathbb{R}^3$, $ i \in \mathbb{N}_1^n $. As shown in~\cite[Corollary 3]{wang2021leading}, this output contains the least information to guarantee the observability of each CF-LCC system. In addition, it ensures that the reference signal $\epsilon_{i+1}$, which is exactly the velocity error of HDV $m_i^{(i)}$, can be acquired by the subsystem $i+1$. Based on this lowest connectivity requirement, the cooperative \method{DeeP-LCC} problem~\eqref{Eq:CooperativeDeePLCCLinearized} can still be solved for control input design.
\end{remark}}

\subsection{Relationship between Cooperative and Centralized Formulations}

We proceed to analyze the relationship between the cooperative formulation~\eqref{Eq:CooperativeDeePLCCLinearized} and the centralized formulation~\eqref{Eq:CentralizedDeePLCCLinearized} in the linearized case with noise-free data. The following assumption is needed on the consistency between the data collected from the two formulations. The result is summarized in Theorem~\ref{Theorem:CostFunction}, and the proof can be found in~\ref{Appendix:Proof}.

\begin{assumption}[Data consistency] \label{Assumption:SameTrajectory}
A same trajectory sequence of mixed traffic flow is under consideration for centralized \method{DeeP-LCC} and cooperative \method{DeeP-LCC} in offline data collection. In other words, an appropriate partition of the centralized data sequence in~\eqref{Eq:CollectedData} leads to the local data sequence in~\eqref{Eq:CollectedDataSubsystem}.
\end{assumption}

\begin{theorem} \label{Theorem:CostFunction}
{Let Assumption~\ref{Assumption:SameTrajectory} hold and $T_i = T,\;i \in \mathbb{N}_1^n$. Given time $t$, the same LTI mixed traffic system with no noise, and a same past trajectory before time $t$,} denote $(g^*,u^*,y^*)$ as the optimal solution of centralized linear \method{DeeP-LCC} problem~\eqref{Eq:CentralizedDeePLCCLinearized}, and $(g_i^*,u_i^*,y_i^*),\;i \in \mathbb{N}_1^n$ as the optimal solution of cooperative linear \method{DeeP-LCC} problem~\eqref{Eq:CooperativeDeePLCCLinearized}. Then, it holds that
\begin{equation} \label{Eq:CostRelationship}
    \sum_{i=1}^n V_i(u_i^*,y_i^*) = V(u^*,y^*).
\end{equation}
\end{theorem}

{Theorem~\ref{Theorem:CostFunction} reveals that for the noise-free and LTI mixed traffic system with the same offline pre-collected data and online past trajectories, cooperative linear \method{DeeP-LCC}~\eqref{Eq:CooperativeDeePLCCLinearized} could achieve the identical optimal system-wide stabilizing performance~\eqref{Eq:CostDefinition} compared to centralized linear \method{DeeP-LCC}~\eqref{Eq:CentralizedDeePLCCLinearized}. }
Besides this equivalent optimal behavior, cooperative \method{DeeP-LCC}~\eqref{Eq:CooperativeDeePLCCLinearized} also allows for distributed optimization, which will be detailed in the next section. In addition, fewer pre-collected data points are needed for cooperative \method{DeeP-LCC}, which are collected and stored locally in each CAV. Precisely, to satisfy Assumption~\ref{Assumption:PersistentExcitation} for centralized \method{DeeP-LCC}, the lower bound for the length of centralized data~\eqref{Eq:CollectedData} is given by
\begin{equation}
\label{Eq:CentralizedDataLowerBound}
    T \geq (n+1)(T_{\mathrm{ini}}+N+2m+2n)-1,
\end{equation}
while in Assumption~\ref{Assumption:PersistentExcitationSubsystem} for cooperative \method{DeeP-LCC}, the minimum length for local data~\eqref{Eq:CollectedDataSubsystem} is reduced to 
\begin{equation}
\label{Eq:DistributedDataLowerBound}
    T_i \geq 2(T_{\mathrm{ini}}+N+2m_i+2)-1,\;\; i\in\mathbb{N}_1^n.
\end{equation}

\subsection{Final Design for Cooperative \method{DeeP-LCC}}

Considering a practical mixed traffic setup with nonlinear and non-deterministic behavior and noise-corrupted data, we introduce the following regularized cost function motivated by the design~\eqref{Eq:CentralizedCostFunction} in centralized \method{DeeP-LCC}
\begin{equation}
  \label{Eq:CostFunctionCooperativeDeePLCC}
     J_i(g_i,u_i,y_i,\sigma_{y_i}) = V_i(y_i,u_i)+\lambda_{g_i} \left\|g_i\right\|_2^2+\lambda_{y_i} \left\|\sigma_{y_i}\right\|_2^2.
 \end{equation}
where $\sigma_{y_i} \in \mathbb{R}^{(1+m_i)T_\mathrm{ini}},\; i\in\mathbb{N}_1^n$ are slack variables to ensure feasibility, and $\lambda_{y_i},\lambda_{g_i}>0,\; i\in\mathbb{N}_1^n$ are  weight coefficients for regularization. Then, similarly to the final centralized \method{DeeP-LCC} formulation~\eqref{Eq:CentralizedDeePLCC}, the following regularized version of cooperative \method{DeeP-LCC} is provided

\vspace{0.2em}
\noindent\textbf{Cooperative \method{DeeP-LCC}:}
\begin{subequations} \label{Eq:CooperativeDeePLCC}
 \begin{align}
 \min_{ \substack{g_i,u_i,y_i,\sigma_{y_i}  \\
      i\in \mathbb{N}_1^{n}} } \quad & \sum_{i=1}^n J_i(g_i,u_i,y_i,\sigma_{y_i}) \label{Eq:CooperativeDeePLCCCost}\\
 \st \quad & \begin{bmatrix}
 U_{i,\mathrm{p}} \\ E_{i,\mathrm{p}}\\Y_{i,\mathrm{p}} \\ U_{i,\mathrm{f}} \\ E_{i,\mathrm{f}}\\ Y_{i,\mathrm{f}}
 \end{bmatrix}g_i=
 \begin{bmatrix}
 u_{i,\mathrm{ini}} \\ \epsilon_{i,\mathrm{ini}}\\ y_{i,\mathrm{ini}} \\ u_i \\\epsilon_i \\ y_i
 \end{bmatrix}+\begin{bmatrix}
 0\\0\\ \sigma_{y_i} \\0 \\0 \\0
 \end{bmatrix} ,\;\;  i \in \mathbb{N}_1^{n}, \label{Eq:LocalNonlinearBehavior}\\ &\epsilon_{1} = \hat{\epsilon},\\
 &\epsilon_{i+1} = K_i y_i ,\;\;i \in \mathbb{N}_1^{n-1},
 \\ & u_i\in \mathcal{U}_i,y_i\in \mathcal{Y}_i ,\;\;i \in \mathbb{N}_1^{n}.
 \end{align}
\end{subequations}
This is the finalized formulation for cooperative \method{DeeP-LCC} that could be applied to CAV control in practical mixed traffic flow. As shown in the numerical simulations in~\cite{wang2022deeplcc} and the field experiments~\cite{wang2022implementation}, the estimation of the future velocity error of the head vehicle can be chosen as
\begin{equation}
    \hat{\epsilon}  = \mathbb{0}_N,
\end{equation}
which, combined with a moderate estimation of equilibrium velocity, contributes to satisfactory traffic performance.
For the input/output constraints, given the actuation limit for vehicle longitudinal dynamics, the input constraints $u_i \in \mathcal{U}_i$ are designed as
\begin{equation} \label{Eq:InputConstraint}
a_\mathrm{min} \leq u_i \leq a_\mathrm{max},\;\;i\in \mathbb{N}_1^n,
\end{equation} 
where $	a_\mathrm{min}$ and $a_\mathrm{max}$ denote the minimum and the maximum acceleration, respectively. In addition, an upper bound $\tilde{s}_\mathrm{max}$ and a lower bound $\tilde{s}_\mathrm{min}$ are imposed on the spacing error of each CAV, given by
\begin{equation} \label{Eq:SpacingConstraint}
	\tilde{s}_\mathrm{min} \leq \tilde{s}_{i} \leq \tilde{s}_\mathrm{max},\;\;i\in \mathbb{N}_1^n,
\end{equation}
for CAV collision-free guarantees and a practical consideration that the CAV would not leave an extremely large spacing from the preceding vehicle. Note that $\tilde{s}_\mathrm{min}=s_\mathrm{min}-s^*,\tilde{s}_\mathrm{max}=s_\mathrm{max}-s^*$ with $s_\mathrm{min},s_\mathrm{max}$ denoting the minimum and maximum spacing respectively. This constraint~\eqref{Eq:SpacingConstraint} on the spacing errors is captured by the output constraint, formulated as
\begin{equation} \label{Eq:OutputConstraint}
	\tilde{s}_\mathrm{min} \leq P_i y_i \leq \tilde{s}_\mathrm{max} ,\;\;i \in \mathbb{N}_1^n,
\end{equation}
with
\begin{equation*}
    P_i = I_{N} \otimes \begin{bmatrix}\mathbb{0}_{1\times (m_i+1)} & 1  \end{bmatrix} ,\;\; i \in \mathbb{N}_1^n.
\end{equation*}

{
\begin{remark}
    Based on the design for the cost function~\eqref{Eq:CostFunctionCooperativeDeePLCC} and the input/output constraints~\eqref{Eq:InputConstraint}-\eqref{Eq:OutputConstraint}, the final cooperative \method{DeeP-LCC} problem~\eqref{Eq:CooperativeDeePLCC} is a quadratic programming (QP), which is convex and can also be directly solved by a QP solver in a centralized manner just like the centralized \method{DeeP-LCC} problem~\eqref{Eq:CentralizedDeePLCC}. Different from the centralized version, however, the cooperative formulation requires local traffic data~\eqref{Eq:CollectedDataSubsystem}, which is easier to be obtained, and further allows for distributed optimization, leading to efficient computation for large-scale traffic flow. This will be detailed in the following section. Additionally, according to the equivalence property in Theorem~\ref{Theorem:CostFunction}, given LTI mixed traffic systems with noise-free data, the feasibility of the cooperative formulation is consistent with that of the centralized one, which is essentially an adapted formulation from standard DeePC~\cite{coulson2019data}. We refer the interested readers to~\cite[Proposition 1]{berberich2020data} for theoretical guarantees on recursive feasibility of DeePC under terminal constraints for stabilizing the system and bounded assumptions on the slacking variable and the noise level.
\end{remark}
}

\section{Distributed Implementation via ADMM for Cooperative \method{DeeP-LCC}}
\label{Sec:5}

In the cooperative \method{DeeP-LCC} problem~\eqref{Eq:CooperativeDeePLCC}, the coupling constraint arises from the interaction, \ie, information exchange, between the neighbouring CF-LCC subsystems. In this section, we present a tailored ADMM based distributed algorithm to solve this problem.

\subsection{Review of ADMM}

We first review the basics of the ADMM algorithm~\cite{boyd2011distributed}. By slight abuse of notations, the symbols are used for corresponding representations only in this subsection. Given two groups of decision variables $x \in \mathbb{R}^n,y\in \mathbb{R}^m$ and convex functions $f$ and $h$,
ADMM aims at solving a composite optimization problem of the form
\begin{subequations} \label{Eq:GeneralOptimization}
\begin{align}
\min_{x,y} \quad & f(x)+h(y)\\
\st \quad & Ax+By = c,
\end{align}
\end{subequations}
where $A \in \mathbb{R}^{p \times n},B \in \mathbb{R}^{p \times m},c\in \mathbb{R}^p$. The augmented Lagrangian of problem~\eqref{Eq:GeneralOptimization} is defined as
\begin{equation}
    L_\rho (x, y, \mu)   =   f(x) + h(y) +\mu^\top  (Ax + By -c) + \frac{\rho}{2} \left\| Ax  +  By  - c \right\|_2^2,
\end{equation}
where $\rho>0$ denotes a penalty parameter. Then, using ADMM to solve~\eqref{Eq:GeneralOptimization} yields the following iterations:
\begin{subequations}
    \begin{align}
    x^+ &= \argmin_x L_\rho(x,y,\mu) \\
    y^+ &= \argmin_y L_\rho(x^+,y,\mu) \\
    \mu^+ &= \mu + \rho(Ax^+ + By^+ -c), 
    \end{align}
\end{subequations}
where $x^+,y^+,\mu^+$ denotes the update of iterate $x,y,\mu$. One can clearly observe that ADMM is a primal-dual optimization algorithm, i.e., it consists of an $x$-minimization step and a $y$-minimization step for primal update, and a dual variable update for $\mu$. As discussed in~\cite{boyd2011distributed}, ADMM guarantees convergence for convex optimizations under mild conditions, and in practice only a few tens of iterations are needed for modest accuracy.

\subsection{Decomposable Formulation of Cooperative \method{DeeP-LCC}}

{To apply ADMM, we need to reformulate the cooperative \method{DeeP-LCC} problem~\eqref{Eq:CooperativeDeePLCC} with two groups of decision variables and separated cost functions as shown in~\eqref{Eq:GeneralOptimization}.} According to the data-centric behavior representation in~\eqref{Eq:LocalNonlinearBehavior}, we have the following equality constraint for $g_i$
\begin{equation}
    \begin{bmatrix}
    U_{i,\mathrm{p}} \\ E_{i,\mathrm{p}}
    \end{bmatrix} g_i = \begin{bmatrix}
    u_{i,\mathrm{ini}} \\ \epsilon_{i,\mathrm{ini}}
    \end{bmatrix},\;\; i\in\mathbb{N}_1^n,
\end{equation}
and the other variables can be represented by ($i\in\mathbb{N}_1^n$)
\begin{equation*}
    \sigma_{y_i} = Y_{i,\mathrm{p}}g_i - y_{i,\mathrm{ini}},\quad     u_i = U_{i,\mathrm{f}}g_i, \quad \epsilon_i = E_{i,\mathrm{f}}g_i, \quad y_i = Y_{i,\mathrm{f}}g_i, \quad \tilde{s}_{i} =  P_i Y_{i,\mathrm{f}}g_i.
\end{equation*}
Regarding $g_i$ as the only decision variable in the cost function $J_i(g_i,u_i,y_i,\sigma_{y_i})$ in~\eqref{Eq:CooperativeDeePLCCCost}, Problem~\eqref{Eq:CooperativeDeePLCC} can be converted to
\begin{subequations} 
\label{Eq:DistributedSimplifiedG}
    \begin{align}
 \min_{\substack{g_i  \\
      i\in \mathbb{N}_1^{n}}}
 \; & \sum_{i=1}^n f_i(g_i)\\
 \st \; & \begin{bmatrix}
    U_{i,\mathrm{p}} \\ E_{i,\mathrm{p}}
    \end{bmatrix} g_i  =  \begin{bmatrix}
    u_{i,\mathrm{ini}} \\ \epsilon_{i,\mathrm{ini}}
    \end{bmatrix} ,\;\;i \in \mathbb{N}_1^{n},\label{Eq:DistributedSimplifiedGconstraint1}\\
 &E_{1,\mathrm{f}}g_1  =  \mathbb{0}_N, \label{Eq:DistributedSimplifiedGconstraint2}\\
    &E_{i+1,\mathrm{f}}g_{i+1}   =   K_i Y_{i,\mathrm{f}}g_i ,\;\;i \in \mathbb{N}_1^{n-1}, \label{Eq:DistributedSimplifiedGCoupling}\\
    & P_i Y_{i,\mathrm{f}}g_i = \tilde{s}_i ,\;\;  i \in \mathbb{N}_1^{n},\label{Eq:DistributedGS}\\
    & U_{i,\mathrm{f}} g_i = u_i, \;\;  i \in \mathbb{N}_1^{n}, \label{Eq:DistributedGU} \\
 &\tilde{s}_\mathrm{min} \leq \tilde{s}_i \leq \tilde{s}_\mathrm{max} ,\;\;i \in \mathbb{N}_1^{n}, \label{Eq:DistributedSimplifiedS}\\
 & a_\mathrm{min} \leq u_i \leq a_\mathrm{max},\;\;i \in \mathbb{N}_1^{n}, \label{Eq:DistributedSimplifiedU}
 \end{align}
\end{subequations}
where
\begin{equation*}
    f_i(g_i) := g_i^\top Q_{g_i} g_i + 2 c_{g_i}^\top g_i + \lambda_{y_i} y_{i,\mathrm{ini}}^\top y_{i,\mathrm{ini}},
\end{equation*}
with
$$
Q_{g_i} = Y_{i,\mathrm{f}}^\top Q_i Y_{i,\mathrm{f}} + U_{i,\mathrm{f}}^\top R_i U_{i,\mathrm{f}} + \lambda_{g_i} I + \lambda_{y_i} Y_{i,\mathrm{p}}^\top  Y_{i,\mathrm{p} }, \quad 
c_{g_i}  = -\lambda_{y_i} Y_{i,\mathrm{p}}^\top y_{i,\mathrm{ini}}. 
$$

We proceed to show how to further simplify and decompose the optimization problem~\eqref{Eq:DistributedSimplifiedG}, by treating $g_i$ as one group of variables and establishing the other one. The equality constraints for $g_i$ in~\eqref{Eq:DistributedSimplifiedGconstraint1} and~\eqref{Eq:DistributedSimplifiedGconstraint2} are incorporated into the domain of function $f_i(g_i)$, which is given by
\begin{equation}
    \mathrm{\textbf{Dom}}(f_i)  = \left\{g_i \vert A_{g_i}g_i=b_{g_i} \right\},
\end{equation}
where
\[
A_{g_1} = \begin{bmatrix}
U_{1,\mathrm{p}} \\ E_{1,\mathrm{p}} \\ E_{1,\mathrm{f}}
\end{bmatrix},b_{g_1} = \begin{bmatrix}
u_{1,\mathrm{ini}} \\ \epsilon_{1,\mathrm{ini}} \\ \mathbb{0}_N
\end{bmatrix}; \quad
A_{g_i} = \begin{bmatrix}
U_{i,\mathrm{p}} \\ E_{i,\mathrm{p}}
\end{bmatrix},b_{g_i} = \begin{bmatrix}
u_{i,\mathrm{ini}} \\ \epsilon_{i,\mathrm{ini}}
\end{bmatrix},\;\;i\in \mathbb{N}_2^{n}.
\]

{The constraint~\eqref{Eq:DistributedSimplifiedGCoupling} depicts the coupling relationship between $g_i$ and $g_{i+1}$, which is not desired for ADMM. To decompose it, we introduce
\begin{equation} \label{Eq:gzcoupling1}
    g_i = z_i ,\;\;i \in \mathbb{N}_1^{n},
\end{equation}
and then the coupling constraint can be converted to
\begin{equation} \label{Eq:gzcoupling2}
    E_{i+1,\mathrm{f}}g_{i+1}  = K_i Y_{i,\mathrm{f}}z_i,\;\;i \in \mathbb{N}_1^{n-1}.
\end{equation}
The physical interpretation of this newly introduced variable $z_i$ is that it represents the assumed value of $g_i$ for subsystem $i+1$. Combining~\eqref{Eq:gzcoupling1} and~\eqref{Eq:gzcoupling2} leads to the equality constraint for the $g_i$ group and $z_i$ group variables. }

The equality constraints~\eqref{Eq:DistributedGS} and~\eqref{Eq:DistributedGU} hold for the $g_i$ group and $\tilde{s}_i,u_i$ group variables, respectively, which are already decomposed. However, the existence of the inequality constraints~\eqref{Eq:DistributedSimplifiedS} and~\eqref{Eq:DistributedSimplifiedU} makes it time-consuming to solve the minimization problem for the augmented Lagrangian at each iteration step of ADMM, which requires a quadratic programming solver for numerical computation. To address this issue, we capture the inequality constraints~\eqref{Eq:DistributedSimplifiedS} and~\eqref{Eq:DistributedSimplifiedU} by the indicator functions\footnote{The indicator function $\mathcal{I}(x)$ over the set $\mathcal{C}$ is defined as $\mathcal{I}(x)=0$ for $x\in \mathcal{C}$ and $\mathcal{I}(x)=+\infty$ otherwise.} contained in the objective function. Precisely, we define
\begin{equation}
    h_i(z_i,\tilde{s}_i,u_i) = \mathcal{I}_s(\tilde{s}_i) + \mathcal{I}_u(u_i),
\end{equation}
where $\mathcal{I}_s(s_i),\mathcal{I}_u(u_i)$ are the indicator functions over the sets
$$
\mathcal{C}_s = \{\tilde{s}\in\mathbb{R}^N \vert \tilde{s}_\mathrm{min} \leq \tilde{s} \leq \tilde{s}_\mathrm{max}\}, \;
\mathcal{C}_u = \{u\in\mathbb{R}^N \vert a_\mathrm{min} \leq u \leq a_\mathrm{max}\},
$$
\color{black}
respectively. 

Now, problem~\eqref{Eq:DistributedSimplifiedG} can be rewritten as the following decomposable formulation. 

\vspace{0.2em}
{
\noindent\textbf{Decomposable Cooperative \method{DeeP-LCC}:}
\begin{subequations} \label{Eq:ADMMFinalProblem}
    \begin{align}
 \min_{\substack{g_i,z_i,\tilde{s}_i,u_i \\
      i\in \mathbb{N}_1^{n}}}
 \; & \sum_{i=1}^n \left(f_i(g_i) +  h_i(z_i,\tilde{s}_i,u_i)\right) \\
 \st \; & g_i = z_i,\;\;i \in \mathbb{N}_1^{n},\label{Eq:ADMMFinalProblemGZequal}\\
    & E_{i+1,\mathrm{f}}g_{i+1}   =   K_i Y_{i,\mathrm{f}}z_i,\;\;i \in \mathbb{N}_1^{n-1},   \label{Eq:ADMMFinalProblemGZconnected}\\
    & P_i Y_{i,\mathrm{f}}g_i = \tilde{s}_i ,\;\;  i \in \mathbb{N}_1^{n},\label{Eq:ADMMFinalProblemS}\\
    & U_{i,\mathrm{f}} g_i = u_i, \;\;  i \in \mathbb{N}_1^{n}, \label{Eq:ADMMFinalProblemU}
 \end{align}
\end{subequations}
{with two sets of decision variables $(g_1,\ldots,g_n)$ and $(z_1,\ldots,z_n,\tilde{s}_1,\ldots,\tilde{s}_n,u_1,\ldots,u_n)$, corresponding to the decision variables $x$ and $y$ respectively in the standard form~\eqref{Eq:GeneralOptimization}, as well as separated objective functions $\sum_{i=1}^n f_i(g_i),\;g_i\in \mathrm{\textbf{Dom}}(f_i)$ and $\sum_{i=1}^n h_i(z_i,\tilde{s}_i,u_i)$.} In the following section, we will elaborate how to tailor ADMM to solve~\eqref{Eq:ADMMFinalProblem}.}

\subsection{ Distributed Algorithm for Cooperative \method{DeeP-LCC}}
Now we are ready to present the distributed optimization algorithm based on ADMM to solve Problem~\eqref{Eq:ADMMFinalProblem}. In particular, we aim to design a distributed optimization algorithm, where each CAV serves as the computing unit and communication node of each CF-LCC subsystem. The algorithm is presented as follows.

First, the augmented Lagrangian of Problem~\eqref{Eq:ADMMFinalProblem} is given by
\begin{equation} \label{Eq:Lagrangian}
    L =  \sum_{i=1}^n L^{(1)}(g_i,z_i) + \sum_{i=1}^{n-1} L^{(2)}(g_{i+1},z_i) 
     + \sum_{i=1}^n L^{(3)}(g_i,\tilde{s}_i) + \sum_{i=1}^n L^{(4)}(g_i,u_i),
    \end{equation}
where 
\begin{subequations} 
\begin{align} 
    L^{(1)} (\cdot) &  =   f_i(g_i) + \mu_i^\top(g_i-z_i) + \frac{\rho}{2}\left\| g_i-z_i \right\|_2^2 \\
    L^{(2)} (\cdot) &  =   \eta_i^\top (E_{i+1,\mathrm{f}}g_{i+1}    -    K_i Y_{i,\mathrm{f}}z_i)  +  \frac{\rho}{2}   \left\| E_{i+1,\mathrm{f}}g_{i+1}    -    K_i Y_{i,\mathrm{f}}z_i \right\|_2^2 \\
    L^{(3)} (\cdot) &  =  \mathcal{I}_s(s_i) + \phi_i ^ \top (s_i - P_i Y_{i,\mathrm{f}} g_i) + \frac{\rho}{2}\left\| s_i - P_i Y_{i,\mathrm{f}} g_i \right\|_2^2 \\
    L^{(4)} (\cdot) &  =  \mathcal{I}_u(u_i)  + \theta_i^ \top (u_i - U_{i,\mathrm{f}}g_i) + \frac{\rho}{2}\left\| u_i - U_{i,\mathrm{f}}g_i \right\|_2^2,
\end{align}
\end{subequations}
with $\mu_i,\eta_i,\phi_i,\theta_i$ denoting the dual variables for the equality constraints~\eqref{Eq:ADMMFinalProblemGZequal}--\eqref{Eq:ADMMFinalProblemU} respectively. 

Then, the algorithm consists of the following three steps.

\vspace{0.2em}
\noindent \textbf{Step 1: Parallel Optimization to Update $g_i$:} 
We obtain $g_i^+$ by minimizing $L$ in~\eqref{Eq:Lagrangian} over $g_i$
\begin{equation} \label{Eq:GUpdatePrimal}
\begin{aligned}
        g_i^+ = & \argmin_{g_i \in \mathrm{\bold{Dom}}(f_i)} 
        L^{(1)}(g_i,z_i) + L^{(3)}(g_i,\tilde{s}_i) + L^{(4)}(g_i,u_i) + L^{(2)}(g_i,z_{i-1}) \bigg|_{i \in \mathbb{N}_2^{n}}    \\
        = & \argmin_{g_i \in \mathrm{\bold{Dom}}(f_i)} g_i^\top H_{g_i} g_i  + 2q_{g_i}^\top g_i,
\end{aligned}
\end{equation}
where the coefficients $H_{g_i},q_{g_i}$ are given by
$$
\begin{aligned}
    H_{g_i}   = & Y_{i,\mathrm{f}}^\top Q_i Y_{i,\mathrm{f}} + U_{i,\mathrm{f}}^\top R_i U_{i,\mathrm{f}} + \lambda_{g_i} I + \lambda_{y_i} Y_{i,\mathrm{p}}^\top  Y_{i,\mathrm{p} }   +  \frac{\rho}{2} (I  +   Y_{i,\mathrm{f}}^\top P_i ^\top P_i Y_{i,\mathrm{f}}  +  U_{i,\mathrm{f}}^\top U_{i,\mathrm{f}} )  +   \left( \frac{\rho}{2}E_{i,\mathrm{f}}^\top E_{i,\mathrm{f}} \right)    \bigg|_{i\in \mathbb{N}_2^{n}} 
\\
q_{g_i}    = & \frac{1}{2} \left(\mu_iI  -  \rho z_i  -   Y_{i,\mathrm{f}}^\top P_i ^ \top \phi_i  -  U_{i,\mathrm{f}}^\top \theta_i  -  \rho Y_{i,\mathrm{f}}^\top P_i ^ \top \tilde{s}_i  -  \rho U_{i,\mathrm{f}}^\top  u_i\right)   -   \lambda_{y_i}  Y_{i,\mathrm{p}}^\top y_{i,\mathrm{ini}}  +   \frac{1}{2} E_{i,\mathrm{f}}^\top\bar{\eta}_{i-1}   {\bigg|_{ i\in \mathbb{N}_2^{n}}},
\end{aligned}
$$
with
\begin{equation} \label{Eq:CalculateEta}
    \bar{\eta}_i = \eta_{i} - \rho K_{i} Y_{i,\mathrm{f}}z_{i},\;\;i\in \mathbb{N}_1^{n-1}.
\end{equation}
Since $\mathrm{\bold{Dom}}(f_i)$ is an equality constrained set, the update~\eqref{Eq:GUpdatePrimal} involves solving the following KKT (Karush-Kuhn-Tucker) system
\begin{equation} \label{Eq:KKTsystem}
    \begin{bmatrix}
    H_{g_i} & A_{g_i}^\top \\ A_{g_i} & 0
    \end{bmatrix} \begin{bmatrix}
    g_i^+ \\ \nu
    \end{bmatrix} = \begin{bmatrix}
    -q_{g_i} \\ b_{g_i}
    \end{bmatrix},
\end{equation}
where $\nu$ denotes a Lagrangian multiplier. Since $H_{g_i}$ and $A_{g_i}$ consist of only pre-collected data and pre-determined parameters, the KKT matrix in~\eqref{Eq:KKTsystem} (the left-hand multiplier) is fixed during the entire online control process. Thus, the KKT matrix can be pre-factorized before online predictive control, and $g_i^+$ can be calculated from~\eqref{Eq:KKTsystem} quite efficiently. Precisely, we have
\begin{equation} \label{Eq:GiCalculation}
     \begin{bmatrix}
    g_i^+ \\ \nu
    \end{bmatrix} = - \Psi_i \begin{bmatrix}
    q_{g_i} \\ -b_{g_i}\end{bmatrix},\;\;i\in \mathbb{N}_1^n, 
\end{equation}
where
\[
\Psi_i = \begin{bmatrix}
    H_{g_i} & A_{g_i}^\top \\ A_{g_i} & 0
    \end{bmatrix} ^{-1}
\]
is pre-calculated before the control process.

\noindent \textbf{Step 2: Parallel Optimization to Update $z_i,\tilde{s}_i,u_i$:} 
We obtain $z_i^+,\tilde{s}_i^+,u_i^+$ by minimizing $L$ in~\eqref{Eq:Lagrangian} over $z_i,\tilde{s}_i,u_i$, where $g_i$ adopts the updated value in the previous step. Precisely,
\begin{subequations}\label{Eq:ZSUUpdate}
\begin{align}
z_i^+  = & \argmin_{z_i} 
     L^{(1)}(g_i^+,z_i)  +  L^{(2)}(g_{i+1}^+,z_{i}) \bigg|_{i \in \mathbb{N}_1^{n-1}}  
    =  \argmin_{z_i} z_i^\top H_{z_i} z_i  + 2q_{z_i}^\top z_i
    =  -H_{z_i}^{-1}q_{z_i}
 \label{Eq:ZUpdate}\\ 
    \tilde{s}_i^+ =  &\argmin_{\tilde{s}_i} L^{(3)}(g_i^+,\tilde{s}_i)
    = \argmin_{\tilde{s}_i} \mathcal{I}_s(\tilde{s}_i) + \phi_i ^ \top \tilde{s}_i + \frac{\rho}{2}\left\| \tilde{s}_i - P_i Y_{i,\mathrm{f}} g_i^+ \right\|_2^2 
 =  \Pi_{\mathcal{C}_s} (P_i Y_{i,\mathrm{f}} g_i^+ -  \frac{\phi_i}{\rho}) 
 \label{Eq:SUpdate} \\
    u_i^+ =  &\argmin_{u_i} L^{(4)}(g_i^+,u_i)
    = \argmin_{u_i} \mathcal{I}_u(u_i) + \theta_i ^ \top u_i + \frac{\rho}{2}\left\| u_i - U_{i,\mathrm{f}}g_i^+ \right\|_2^2 
 = \Pi_{\mathcal{C}_u} ( U_{i,\mathrm{f}} g_i^+ -  \frac{\theta_i}{\rho}) , \label{Eq:UUpdate}
\end{align}
\end{subequations}
where the coefficients in~\eqref{Eq:ZUpdate} are given by
$$
H_{z_i}   =   \frac{\rho}{2} I + \left( \frac{\rho}{2}Y_{i,\mathrm{f}}^\top K_i^\top K_i Y_{i,\mathrm{f}} \right) \bigg|_{i\in \mathbb{N}_1^{n-1}} 
,\quad
q_{z_i}   =   -\frac{\mu_i}{2}  -  \frac{\rho g_i^+}{2}  -  Y_{i,\mathrm{f}}^\top K_i^\top\left(  \frac{ \eta_{i}}{2}  +  \frac{\rho  \bar{\epsilon}_{i+1}^+}{2}  \right)  {\bigg|_{i\in \mathbb{N}_1^{n-1}}},
$$
with
\begin{equation} \label{Eq:CalculateEpsilon}
    \bar{\epsilon}_i^+ = E_{i,\mathrm{f}}g_i^+ ,\;\;i\in \mathbb{N}_2^{n}.
\end{equation}
In~\eqref{Eq:SUpdate} and~\eqref{Eq:UUpdate}, $\Pi_\mathcal{C}$ denotes the projection (in the Euclidean norm) onto the set $\mathcal{C}$. Note that similarly to $\Psi_i$ in~\eqref{Eq:GiCalculation}, $H_{z_i}^{-1}$ can be pre-calculated before the online control process. It is also worth noting that $\mathcal{C}_s$ and $\mathcal{C}_u$ are simple box constrained sets with upper and lower bounds on each entry of the vector, and thus it is trivial to calculate the projections~\eqref{Eq:SUpdate} and~\eqref{Eq:UUpdate}.

\vspace{0.2em}
\noindent \textbf{Step 3: Update Dual Variables:} We update the dual variables $\mu_i^+,\eta_i^+,\phi_i^+,\theta_i^+$ by
\begin{subequations} \label{Eq:DualUpdate}
\begin{align}
\mu_i^+  = & \mu_i  +  \rho (g_i^+-z_i^+),\;\; i \in \mathbb{N}_1^{n}, \\
\eta_i^+   =& \eta_i  +  \rho (\bar{\epsilon}_{i+1}^+   -  K_{i} Y_{i,\mathrm{f}}z_{i}^+),\;\; i \in \mathbb{N}_1^{n - 1}, \\
\phi_i^+   =& \phi_i  +  \rho (\tilde{s}_i^+ - P_i Y_{i,\mathrm{f}} g_i^+),\;\;i \in \mathbb{N}_1^{n}, \\
\theta_i^+   =& \theta_i  +  \rho (u_i^+ - U_{i,\mathrm{f}}g_i^+),\;\; i \in \mathbb{N}_1^{n}.
\end{align}
\end{subequations}

This tailored ADMM algorithm solving cooperative \method{DeeP-LCC}~\eqref{Eq:CooperativeDeePLCC} in a distributed manner is summarized in Algorithm~\ref{Alg:DistributedDeePLCC}, named as distributed \method{DeeP-LCC} algorithm. {Each CAV $i$ locally makes calculations in parallel and exchanges necessary data with its neighbours $i-1$ and $i+1$.} In particular, \emph{this algorithm includes only simple numerical calculations, without any optimization problem to be solved.} The convergence proofs of ADMM are available in~\cite[Appendix~A]{boyd2011distributed} and the references therein. The stopping criterion is presented in~\ref{Appendix:Stopping}. Note that distributed \method{DeeP-LCC} is implemented in a receding horizon manner as time moves forward, similarly to the centralized \method{DeeP-LCC} method~\cite{wang2022deeplcc}. In addition, given time $t$, the optimal values of the variables $g_i,z_i,\tilde{s}_i,u_i,\mu_i,\eta_i,\phi_i,\theta_i$ are utilized as their initial values for the ADMM iterations at time $t+1$, as shown in line 16 at  Algorithm~\ref{Alg:DistributedDeePLCC}.  

\begin{figure*}[tb!]
	\centering
	\includegraphics[width=10cm]{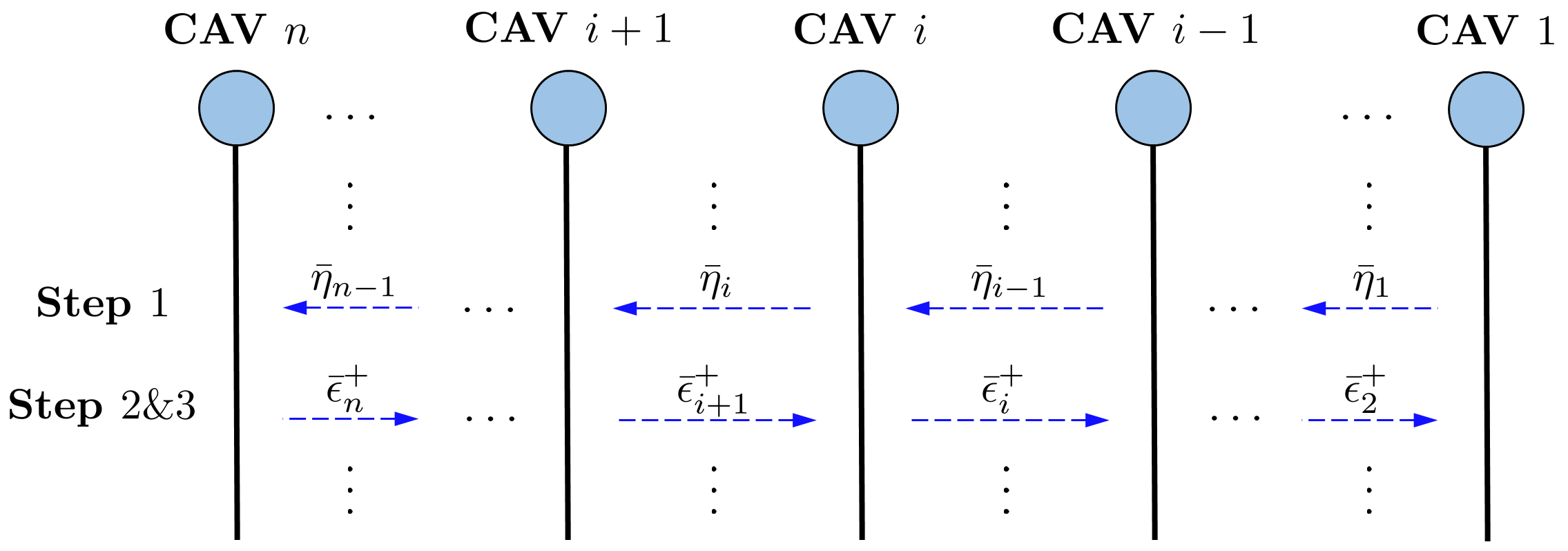}
	\caption{Exchange of computing data in distributed \method{DeeP-LCC}. In each iteration, the values $\bar{\eta}_i,\bar{\epsilon}_i^+$ defined in~\eqref{Eq:CalculateEta} and~\eqref{Eq:CalculateEpsilon} are locally calculated in CAV $i$ and then sent to the corresponding receiver.} 
	\label{Fig:InformationFlow}
\end{figure*}

\begin{algorithm}[t!]
	\caption{Distributed \method{DeeP-LCC}}
	\label{Alg:DistributedDeePLCC}
	\begin{algorithmic}[1]
		\State \textbf{Input:}
		Initial time $t_0$, terminal time $t_f$, data Hankel matrices $U_{i,\mathrm{p}},U_{i,\mathrm{f}},$ $E_{i,\mathrm{p}},E_{i,\mathrm{f}},Y_{i,\mathrm{p}},Y_{i,\mathrm{f}}$;
		\State \textbf{Pre-calculation:} For $i\in \mathbb{N}_1^n$, each CAV $i$ in parallel calculates the value of $\Psi_i$ and $H_{z_i}^{-1}$;
		\State \textbf{Initialization}: For $i\in \mathbb{N}_1^n$, each CAV $i$ initializes $g_i,z_i,\tilde{s}_i,u_i,\mu_i,\eta_i,\phi_i,\theta_i \gets 0$;
		\While{$t_0 \leq t \leq t_f$}
		\State For $i\in \mathbb{N}_1^n$, each CAV $i$ updates past trajectory data $(u_{i,\mathrm{ini}},\epsilon_{i,\mathrm{ini}},y_{i,\mathrm{ini}})$ before $t$;
		\While{stopping criteria~\eqref{Eq:StoppingCriteria} is not satisfied}
		\State For $i\in \mathbb{N}_1^{n-1}$, each CAV $i$ in parallel calculates the following vector signal and sends it to CAV $i+1$
		$$
		\bar{\eta}_i=\eta_{i} - \rho K_{i} Y_{i,\mathrm{f}}z_{i};
		$$ 
		\State For $i\in \mathbb{N}_1^{n}$, each CAV $i$ in parallel calculates the following equation for the value of $g_i^+$
		$$
		\begin{bmatrix}
    g_i^+ \\ \nu
    \end{bmatrix} = - \Psi_i \begin{bmatrix}
    q_{g_i} \\ -b_{g_i}\end{bmatrix};
		$$
		\State For $i\in \mathbb{N}_2^n$, each CAV $i$ in parallel calculates the following vector signal and sends it to CAV $i-1$
		$$
		\bar{\epsilon}_i^+ = E_{i,\mathrm{f}}g_i^+;
		$$ 
		\State For $i\in \mathbb{N}_1^{n}$, each CAV $i$ in parallel calculates 
		$$
		z_i^+=-H_{z_i}^{-1}q_{z_i},\;\;\tilde{s}_i^+=\Pi_{\mathcal{C}_s} (P_i Y_{i,\mathrm{f}} g_i^+ -  \frac{\phi_i}{\rho}),\;\;u_i^+=\Pi_{\mathcal{C}_u} ( U_{i,\mathrm{f}} g_i^+ -  \frac{\theta_i}{\rho});
		$$
		\State For $i\in \mathbb{N}_1^{n}$, each CAV $i$ in parallel obtains
		$$
		\mu_i^+ = \mu_i  +  \rho (g_i^+-z_i^+),\;\;
		\phi_i^+=\phi_i  +  \rho (\tilde{s}_i^+ - P_i Y_{i,\mathrm{f}} g_i^+),\;\;
		\theta_i^+=\theta_i  +  \rho (u_i^+ - U_{i,\mathrm{f}}g_i^+);
		$$
		\hspace{2.7em} meanwhile, for $i \in \mathbb{N}_1^{n-1}$, each CAV $i$  in parallel calculates 
		$$
		\eta_i^+=\eta_i  +  \rho (\bar{\epsilon}_{i+1}^+   -  K_{i} Y_{i,\mathrm{f}}z_{i}^+);
		$$
		\State For $i\in \mathbb{N}_1^{n}$, set $g_i,z_i,\tilde{s}_i,u_i, \mu_i,\eta_i,\phi_i,\theta_i \gets g_i^+,z_i^+,\tilde{s}_i^+,u_i^+,\mu_i^+,\eta_i^+,\phi_i^+,\theta_i^+$ respectively.
		\EndWhile
		\State For $i\in \mathbb{N}_1^n$, denote the final values from the iterations as $g_i^*,z_i^*,\tilde{s}_i^*,u_i^*, \mu_i^*,\eta_i^*,\phi_i^*,\theta_i^* \gets g_i,z_i,\tilde{s}_i,u_i,\mu_i,$ $\eta_i,\phi_i,\theta_i$ respectively. Particularly, $u_i^*$ is expressed by $u_i^*=\col(u_i^*(t),u_i^*(t+1),\ldots,u_i^*(t+N-1))$;
		\State For $i\in \mathbb{N}_1^n$, CAV $i$ applies the input $u_i(t) \gets u_i^*(t)$;
		\State $t \leftarrow t+1$;
		\State For $i\in \mathbb{N}_1^n$, each CAV $i$ sets the initial values for the next round of iterations as $g_i,z_i,\tilde{s}_i,u_i,\mu_i,\eta_i,\phi_i,\theta_i \gets g_i^*,z_i^*,\tilde{s}_i^*,u_i^*, \mu_i^*,\eta_i^*,\phi_i^*,\theta_i^*$ respectively;
		\EndWhile
	\end{algorithmic}
\end{algorithm}

\begin{remark}[Information Flow, Communication Efficiency and Data Privacy]
For distributed \method{DeeP-LCC} implementation, each CAV $i$ serves as the computing unit and communication node for CF-LCC subsystem $i$. Inside the subsystem, it monitors the motion of the following HDVs and the HDV immediately ahead for trajectory data; see the gray solid arrows in Fig.~\ref{Fig:CFLCCSubsystem}. Meanwhile, CAV $i$ exchanges computing data with its neighbouring CAVs $i-1$ and $i+1$; see Fig.~\ref{Fig:InformationFlow} for illustration of data exchange during the iterations. This cross-subsystem information flow topology, represented as blue dashed arrows in Fig.~\ref{Fig:CFLCCSubsystem}, is known as bidirectional topology in pure-CAV platoons~\cite{zheng2016stability}. {It is worth noting that during each iteration, equivalently an information exchange via V2X communications, only two short vectors of computing data $\bar{\eta}_i,\bar{\epsilon}_i^+ \in \mathbb{R}^N$ defined in~\eqref{Eq:CalculateEta} and~\eqref{Eq:CalculateEpsilon} need to be transmitted to the corresponding receiver. These vectors can be computed locally and transmitted efficiently, requiring low communication resources.} Moreover, the pre-collected trajectory data, contained in data Hankel matrices ($U_{i,\mathrm{p}},U_{i,\mathrm{f}},E_{i,\mathrm{p}},E_{i,\mathrm{f}},Y_{i,\mathrm{p}},Y_{i,\mathrm{f}}$), are well preserved in each local subsystem, contributing to data privacy.
\end{remark}
 
{
\begin{remark}[Practical Deployment of Distributed \method{DeeP-LCC}]
\label{Remark:ParcticalDeployment}
Current commercial V2X communication technologies, such as DSRC or IEEE 802.11p~\cite{kenney2011dedicated}, have limited communication frequencies. In the distributed \method{DeeP-LCC} algorithm, multiple information exchanges can occur in one control step due to the iterations. To use commercial V2X equipment for the algorithm, one can run it for only one or two iterations, leading to limited performance degradation when the traffic flow is far from constraint boundaries. Step 1 in the algorithm solves an unconstrained \method{DeeP-LCC} problem, while Step 2 aims to estimate the reference signal and satisfy input/output constraints. Therefore, when the algorithm stops early, the obtained control inputs could allow for satisfactory performance, despite a potential slight mismatch in the reference estimation. The simulations in Section~\ref{Sec:6} will investigate the performance degradation under this approach. Note that this approach is not suitable when the system is close to a collision, but a low-level safety control algorithm like the standard automatic emergency braking system can ensure safety in such cases. Alternatively, 5G V2X technology, which offers higher communication frequencies and smaller delays, can also be utilized instead of commercial V2X technologies. This technique has garnered increasing attention in recent CAV research~\cite{gyawali2020challenges,molina2017lte}. 
\end{remark}
}

\section{Traffic Simulations}
\label{Sec:6}

This section presents two different scales of nonlinear and non-deterministic traffic simulations to validate the performance of distributed \method{DeeP-LCC} in mixed traffic flow\footnote{The algorithm and simulation scripts are available at \url{https://github.com/PREDICT-EPFL/Distributed-DeeP-LCC}.}. 

\subsection{Experimental Setup}

{Motivated by existing research~\cite{jin2017optimal,di2019cooperative,lan2021data}, a noise-corrupted nonlinear OVM model is utilized in the experiments to capture the dynamics of HDVs, given as follows~\cite{bando1995dynamical}}
\begin{equation}
\label{Eq:OVMModel}
    \dot{v}_i(t)=\alpha\left(v_{\mathrm{des}}\left(s_i(t)\right)-v_i(t)\right)+\beta\dot{s}_i(t)+\delta_a,\;\; i \in \mathcal{F},
\end{equation}
where $\alpha, \beta > 0$ represent the sensitivity coefficients, and $v_{\mathrm{des}}(s)$ denotes the spacing-dependent desired velocity of the human driver, given by
\begin{equation*} \label{Eq:OVMDesiredVelocity}
v_{\mathrm{des}}(s)=\begin{cases}
0, &s\le s_{\mathrm{st}};\\
f_v(s) = \frac{v_{\max }}{2}\left(1-\cos (\pi\frac{s-s_{\mathrm{st}}}{s_{\mathrm{go}}-s_{\mathrm{st}}})\right), &s_{\mathrm{st}}<s<s_{\mathrm{go}};\\
v_{\max}, &s\ge s_{\mathrm{go}}.
\end{cases}
\end{equation*}
In~\eqref{Eq:OVMModel}, the acceleration noise signal $\delta_a$ follows a uniform distribution $\delta_a \sim \mathbb{U}[-0.1,0.1]$, where $\mathbb{U}[\cdot]$ denotes the uniform distribution. { A heterogeneous but fixed parameter setup is employed for the HDVs: $\alpha=0.6+\mathbb{U}[-0.2,0.2]$, $\beta=0.9+\mathbb{U}[-0.2,0.2]$, $s_{\mathrm{go}}=35+\mathbb{U}[-5,5]$. The rest of parameters are set as $v_{\max}=30$, $s_{\mathrm{st}}=5$. } { Note that the OVM model is utilized as an example to capture the HDVs' behaviors and show the performance of our method. It is also applicable to the case of other car-following models in the form of~\eqref{Eq:HDVModel}, \eg, the IDM model~\cite{treiber2000congested}.}

For distributed/centralized \method{DeeP-LCC}, the sampling interval is $\Delta t = 0.05 \,\mathrm{s}$. {In offline data collection, we assume that there exists an i.i.d.~signal of $\mathbb{U}[-1,1]$ on the control input signal $u_i$ of each CAV. In this way, the persistent excitation requirement in Assumptions~\ref{Assumption:PersistentExcitation} and~\ref{Assumption:PersistentExcitationSubsystem} is naturally satisfied given a sufficiently large number of data samples.} {Note that at a greater level of system noise, some data pre-processing methods can be applied to the data Hankel matrices in~\eqref{Eq:LocalNonlinearBehavior}, which are constructed from pre-collected data, to improve the accuracy of data-centric behavior representation; see, \eg, the low-rank approximation in~\cite{dorfler2022bridging}.} 

In online control, the time horizons for the future signal sequence and past signal sequence are set to $N=50$, $T_{\mathrm{ini}}=20$, respectively. {In the cost function~\eqref{Eq:CostDefinition}, the weight coefficients are set to $w_v=1,w_s=0.5,w_u=0.1$, corresponding to a balanced consideration for penalizing velocity errors, spacing errors and control inputs~\cite{wang2020controllability,wang2022deeplcc}.} For input/output constraints, the CAV spacing boundaries are set to $s_{\max} = 40, s_{\min} = 5$, and the acceleration limits are set to $a_{\max} = 2, a_{\min} = -5$ (this limit also holds for all the HDVs via saturation). {In our simulations, we assume that the equilibrium velocity for all the vehicles is known as $v^*=15$, and the equilibrium spacing for the CAVs is designed as $s^*=20$. We refer the interested readers to~\cite{wang2022implementation} for one potential approach to practically estimate the traffic equilibrium states.}

For computation, all the experiments are run in MATLAB 2021a. In centralized \method{DeeP-LCC}, the \method{quadprog} solver is utilized to solve~\eqref{Eq:CentralizedDeePLCC} via the interior point method with the optimality tolerances set to $10^{-3}$. In the distributed \method{DeeP-LCC} algorithm, no solvers are needed for computation, and the absolute and relative tolerances for the stopping criterion in~\ref{Appendix:Stopping} are set to $\delta_{\mathrm{abs}}=0.1,\delta_{\mathrm{rel}}=10^{-3}$. {The penalty parameter is chosen as $\rho=1$, motivated by the standard setup in~\cite{boyd2011distributed}. }

\subsection{Moderate-Scale Validation and Comparison}
\label{Sec:Simulation1}

Our first experiment focuses on moderate-scale validation of distributed \method{DeeP-LCC} and makes comparisons with the centralized version. We consider $15$ vehicles in total behind the head vehicle, among which there exist $5$ CAVs ($n=5$), uniformly distributed in mixed traffic flow. Precisely, the CAVs are located at the $1$st, $4$th, $7$th, $10$th, and $13$th vehicle. The  length for pre-collected data is $T=1200$ for centralized \method{DeeP-LCC} and $T_i = 300,\;i\in\mathbb{N}_1^n$ for distributed \method{DeeP-LCC}. Both data lengths are approximately twice of the lower bound in~\eqref{Eq:CentralizedDataLowerBound} and~\eqref{Eq:DistributedDataLowerBound}, respectively. 
In centralized \method{DeeP-LCC}, the weight coefficients for regulated terms are set to $\lambda_g=10,\lambda_y=10000$, while in distributed \method{DeeP-LCC}, we have  $\lambda_{g_i}=2,\lambda_{y_i}=10000,\;i \in \mathbb{N}_1^n$. {Interested readers are referred to~\cite{coulson2019data} for the influence of the choice of these regulated weights, and here the setup is motivated by the previous results of centralized \method{DeeP-LCC}~\cite{wang2022deeplcc,wang2022data}.}

\begin{figure*}[t]
	\centering
	\subfigure[All HDVs]
	{\includegraphics[width=7.5cm]{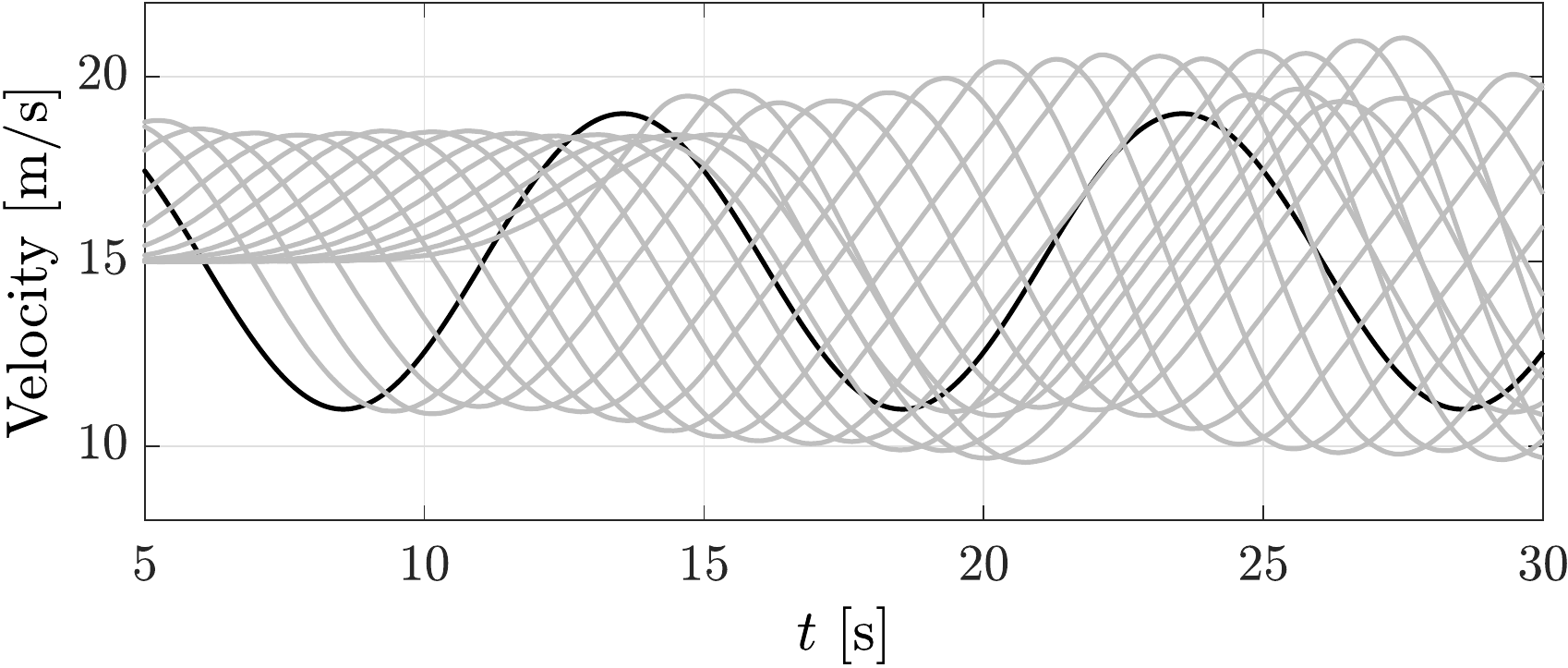}
	\label{Fig:SinusoidPerturbation_HDV}}
	\hspace{0.5cm}
	\subfigure[Distributed \method{DeeP-LCC}]
	{\includegraphics[width=7.5cm]{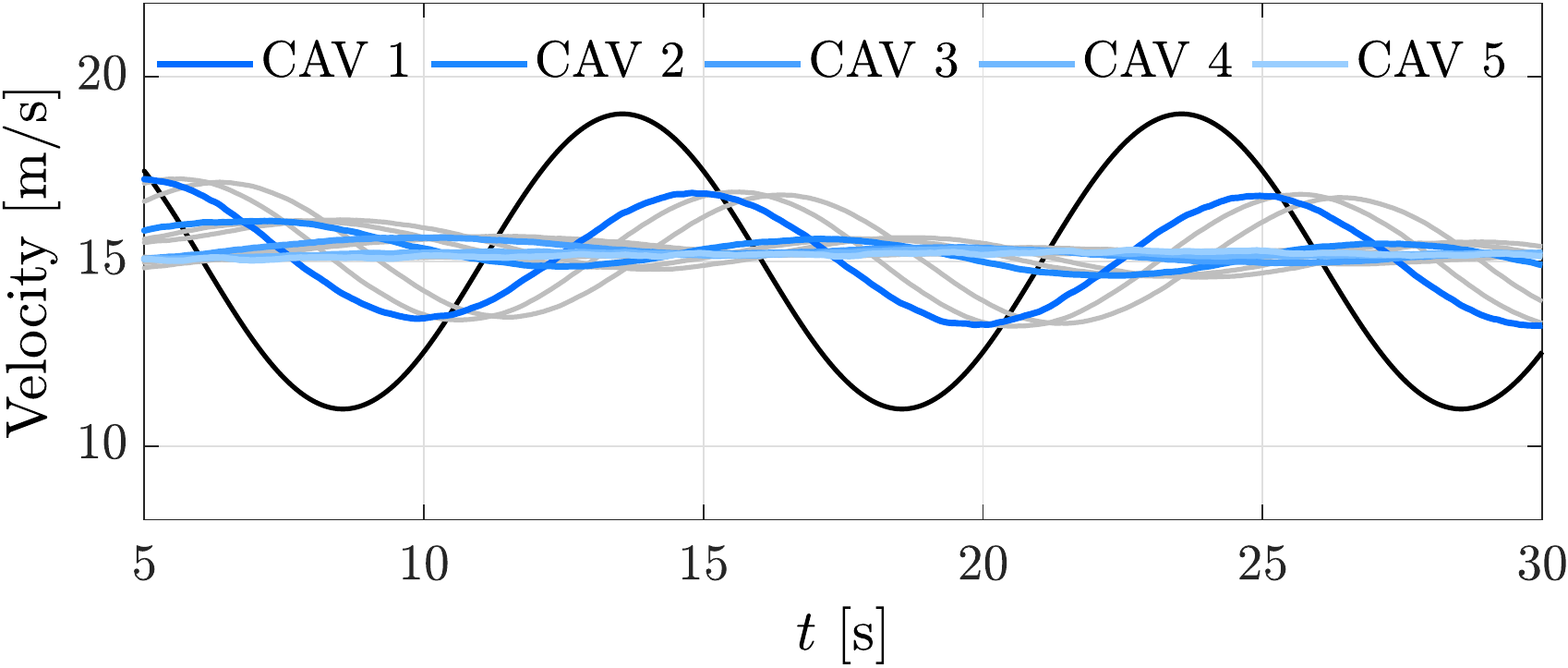}
	\label{Fig:SinusoidPerturbation_CAV}}
	\caption{Velocity profiles in moderate-scale experiments with $15$ vehicles, where a sinusoidal perturbation is imposed on the head vehicle. The black profile represents the head vehicle, while the blue profiles represent the CAVs, with different darkness denoting different positions, and the gray profiles represent the HDVs. (a) All the vehicles are HDVs. (b) The CAVs utilize the distributed \method{DeeP-LCC} controller.}
	\label{Fig:SinusoidPerturbation_VelocityProfile}
\end{figure*}

In this experiment, we assume that the head vehicle is under a sinusoidal perturbation to capture the scenario where the head vehicle is already caught in a traffic wave. Specifically, its velocity is  designed as a sinusoidal profile with a mean value of $15\, \mathrm{m/s}$ (see the black profile in Fig.~\ref{Fig:SinusoidPerturbation_VelocityProfile}).
In the case of all HDVs, the perturbation of the head vehicle is propagating along the vehicle chain, and the amplitude of velocity oscillations is even gradually amplified, as shown in Fig.~\ref{Fig:SinusoidPerturbation_HDV}. In distributed \method{DeeP-LCC}, by contrast, this perturbation is greatly dissipated. In particular, it is observed in Fig.~\ref{Fig:SinusoidPerturbation_CAV} that CAV $2$ is already driving with a relatively smooth velocity and without apparent oscillations, indicating that the traffic wave is almost completely absorbed when it arrives at the second CF-LCC subsystem. This demonstrates the great capability of distributed \method{DeeP-LCC} in reducing traffic instabilities.

\begin{figure*}[t]
	\centering
	\subfigure[Comparison of Real Cost]
	{\includegraphics[width=7.5cm]{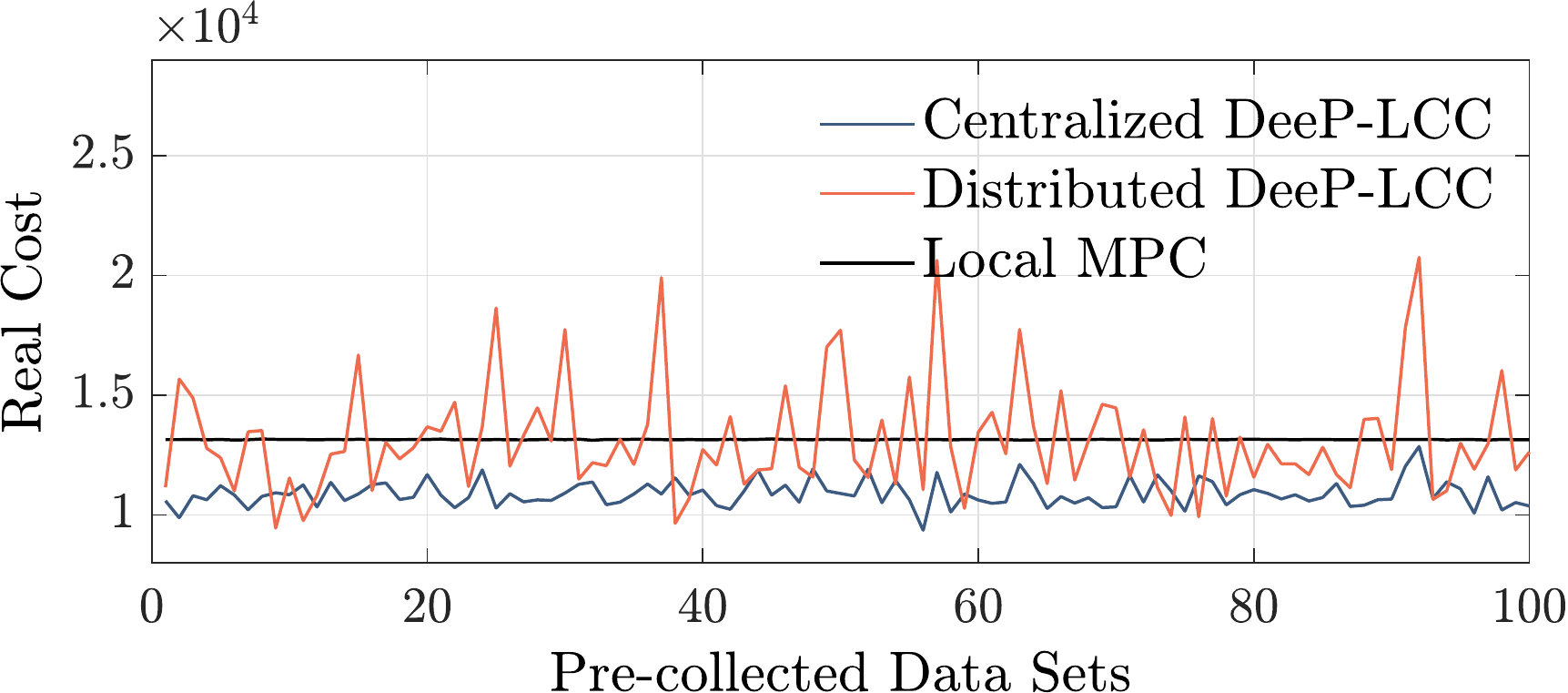}
	\label{Fig:SinusoidPerturbation_RealCost}}
		\hspace{0.5cm}
	\subfigure[Comparison of Computation Time]
	{\includegraphics[width=7.5cm]{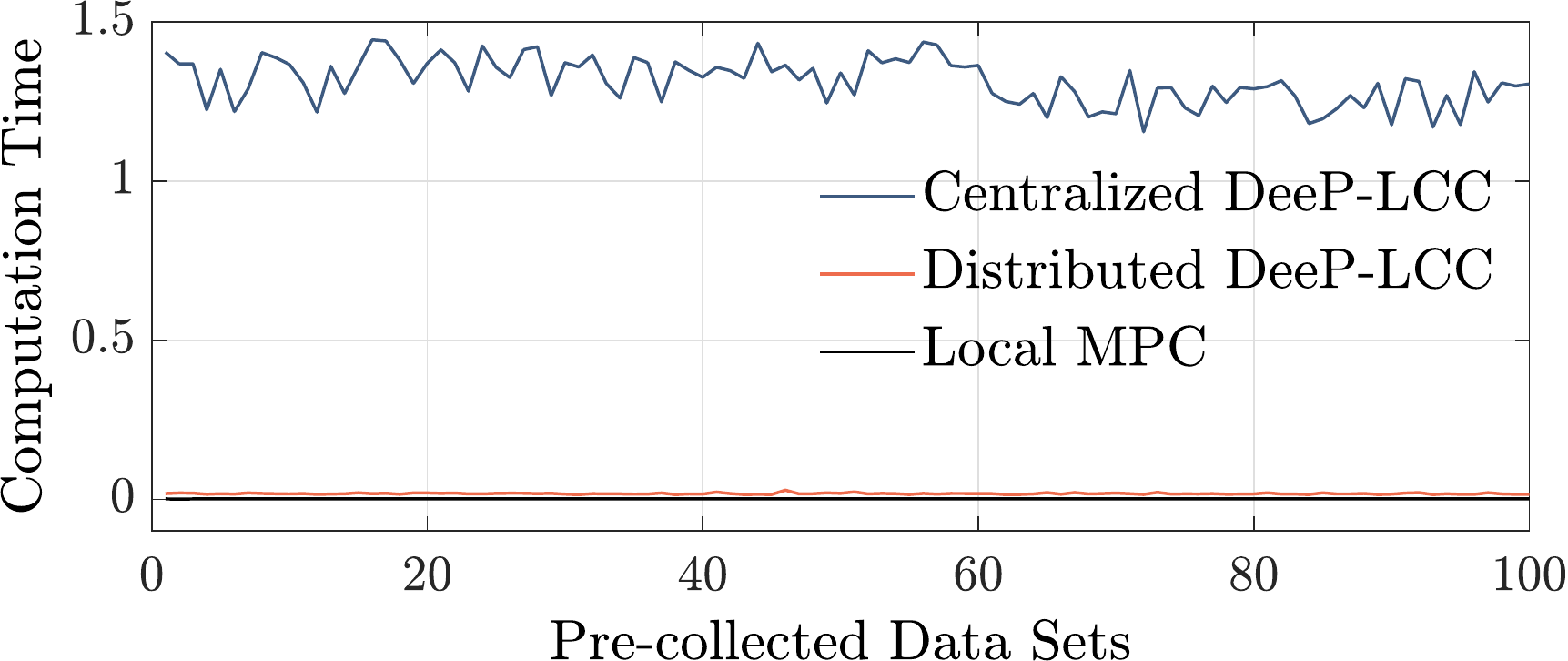}
	\label{Fig:SinusoidPerturbation_ComputationTime}}
	\caption{Comparison of real cost and computation time between distributed \method{DeeP-LCC}, centralized \method{DeeP-LCC} and local MPC in $100$ experiments in the moderate-scale simulation. The average cost for Centralized \method{DeeP-LCC}, distributed \method{DeeP-LCC}, and local MPC is $1.09\times 10^4$, $1.31\times 10^4$, and $1.32\times 10^4$, respectively, while the average computation time for the three methods is $1.27\,\mathrm{s}$, $0.02\,\mathrm{s}$, and $0.001\,\mathrm{s}$, respectively. This experiment is run in MATLAB 2021a with a CPU of Intel Core i7-10700.}
	\label{Fig:SinusoidPerturbation_Statistics}
\end{figure*}

{Since centralized and distributed \method{DeeP-LCC} directly rely on the pre-collected trajectory data~\eqref{Eq:CollectedData} and~\eqref{Eq:CollectedDataSubsystem}, respectively, for online predictive control, the data quality could make a difference on the online predictive control performance. Accordingly, we proceed to make further comparisons between centralized \method{DeeP-LCC} and distributed \method{DeeP-LCC} given $100$ random sets of pre-collected trajectory data. } Fig.~\ref{Fig:SinusoidPerturbation_RealCost} shows the real value of cost function $V$ given by~\eqref{Eq:CostDefinition} at each simulation under centralized or distributed \method{DeeP-LCC}. {As can be clearly observed, centralized \method{DeeP-LCC} achieves a slightly better performance, with distributed \method{DeeP-LCC} losing about $16.79\%$ optimality performance. Note that Theorem~\ref{Theorem:CostFunction} has stated that for a noise-free LTI mixed traffic system, both formulations in the linear setup, \ie,~\eqref{Eq:CentralizedDeePLCCLinearized} and \eqref{Eq:CooperativeDeePLCCLinearized}, could achieve a consistent optimal performance given the same pre-collected trajectory data. The reasons for the performance gap herein between~\eqref{Eq:CentralizedDeePLCC} and~\eqref{Eq:CooperativeDeePLCC} could include: 1) the setup of a nonlinear and noise-corrupted mixed traffic system; 2) the difference in pre-collected trajectory data (distributed \method{DeeP-LCC} utilizes much fewer data samples than centralized \method{DeeP-LCC}); 3) the impact of regularization in~\eqref{Eq:CentralizedDeePLCC} and~\eqref{Eq:CooperativeDeePLCC} with respect to~\eqref{Eq:CentralizedDeePLCCLinearized} and \eqref{Eq:CooperativeDeePLCCLinearized}. } 

{In addition, we also show the performance of local MPC in Fig.~\ref{Fig:SinusoidPerturbation_RealCost}, which utilizes the same local input/output measurement $u_i(t),y_i(t)$ as distributed \method{DeeP-LCC} and employs the accurate model of each CF-LCC subsystem $i$ to design the control inputs for CAV $i$ based on the standard output-feedback MPC framework. This benchmark method is model-based and locally calculated, \ie, there is no cooperation between the different CAVs, and also, it considers a same local control objective $V_i$ in~\eqref{Eq:CostDefinitionSubsystem}. As can be clearly observed, despite the slight performance gap of distributed \method{DeeP-LCC} with respect to centralized \method{DeeP-LCC}, it achieves a comparable performance with local MPC, which, instead, requires prior knowledge of system dynamics. This result shows the potential of distributed \method{DeeP-LCC}, which directly relies on local trajectory data and allows for CAV cooperation in a distributed manner. } 

Furthermore, Fig.~\ref{Fig:SinusoidPerturbation_ComputationTime} demonstrates the real-time computation capability of distributed \method{DeeP-LCC}, costing only a mean computation time of $0.017\,\mathrm{s}$ for each time step, while it takes up to $1.27\,\mathrm{s}$ for solving centralized \method{DeeP-LCC}, which is not tolerable for practical implementation. Thanks to fewer data samples (comparing $T_i=300$ with $T=1200$) and the computationally efficient design of the tailored ADMM algorithm, distributed \method{DeeP-LCC} achieves this dramatic reduction in computation time, while preserving a satisfactory performance for dissipating traffic waves. 

\begin{figure*}[t!]
	\centering
	\includegraphics[width=9.5cm]{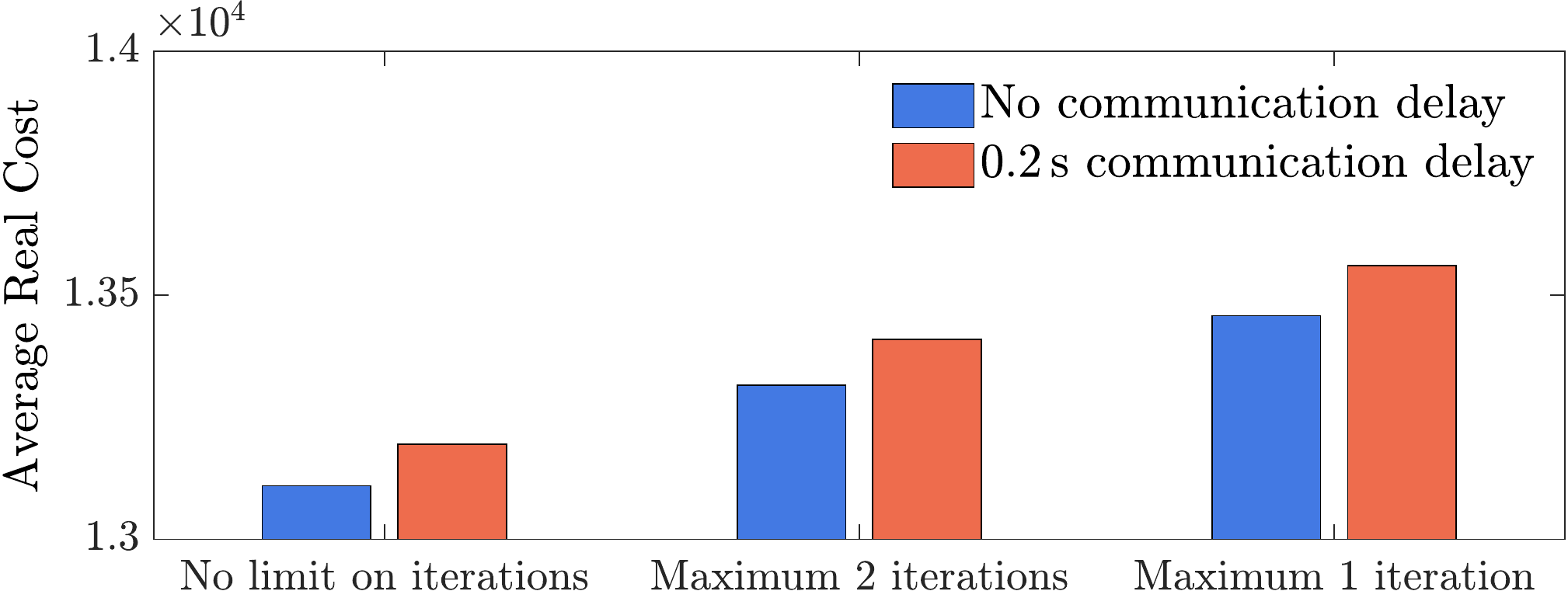}
	\caption{Average real cost of distributed \method{DeeP-LCC} under different imperfect conditions from $100$ experiments in the moderate-scale simulation. The case of no limit on iterations and no communication delay represents the ideal performance of distributed \method{DeeP-LCC}, whose result is consistent with that in Fig.~\ref{Fig:SinusoidPerturbation_Statistics}. The performance degradation percentage with respect to the ideal case is shown in Table~\ref{Tb:PerformanceDegradation}. }
	\label{Fig:SinusoidPerturbation_PerformanceDegradation}
\end{figure*}

\begin{table}[t!]
\footnotesize
	\begin{center}
		\caption{Performance Degradation Percentage of Average Real Cost under Imperfect Conditions}\label{Tb:PerformanceDegradation}
		\begin{threeparttable}
		\setlength{\tabcolsep}{5mm}{
		\begin{tabular}{cccc}
		\toprule
			 & No limit on iterations & Maximum $2$ iterations & Maximum $1$ iteration \\\hline
			 No communication delay & $0$ & $1.57\%$ & $2.66\%$ \\
    $0.2\,\mathrm{s}$ communication delay & $0.65\%$ & $2.28\%$ & $3.44\%$\\
			\bottomrule
		\end{tabular}}
		\end{threeparttable}
	\end{center}
	\vspace{-5mm}
\end{table}

{Finally, we also investigate the performance degradation of distributed \method{DeeP-LCC} under typical imperfect conditions. We consider two issues: (1) an upper bound on the iteration number of the algorithm due to the low communication frequency of commercial V2X equipment, as discussed in Remark~\ref{Remark:ParcticalDeployment}; and (2) communication delay in the V2X communications. We compare the results of the ideal distributed \method{DeeP-LCC} (\ie, without iteration limit and communication delay) with those obtained under different imperfect conditions. The average real cost of different cases from $100$ random experiments is shown in Fig.~\ref{Fig:SinusoidPerturbation_PerformanceDegradation}, and the performance degradation percentage with respect to the ideal case is listed in Table~\ref{Tb:PerformanceDegradation}. Our findings suggest that, even under imperfect conditions, distributed \method{DeeP-LCC} can provide satisfactory traffic smoothing performance, with no more than $3.44\%$ average real cost loss when the algorithm iterates only once or twice or there is a  communication delay of $0.2\,\mathrm{s}$. This demonstrates the robustness of the method against imperfections in the V2X network. However, when the traffic system is close to collisions or there is a large communication delay, significant performance degradation is expected. Therefore, future work should focus on adjusting the proposed algorithm to explicitly address these imperfect conditions. }

\subsection{Large-Scale Experiments}
\label{Sec:Simulation2}

{Our second experiment aims to validate the cooperative control performance of distributed \method{DeeP-LCC} in large-scale mixed traffic flow, where there are $100$ vehicles following behind the head vehicle. Different penetration rates of the CAVs are also under consideration, including $5\%$, $10\%$, and $20\%$. In this experiment, the CAVs are not uniformly distributed among the following vehicles, and there could be different numbers of HDVs in each CF-LCC subsystem. Precisely,  Table~\ref{Tb:ParameterSetupLargeScale} lists the possible choices of the HDV number $m_i$ behind each CAV, and some parameter values of distributed \method{DeeP-LCC} in each penetration rate. Note that the weight coefficients for the cost function are slightly adjusted for different penetration rates.}
The velocity trajectory for the head vehicle is designed as shown in Table~\ref{Tb:HeadVehicleTrajectory}, which takes an emergency brake during the simulations. An instantaneous fuel consumption model 
is employed to calculate the fuel economy of traffic flow~\cite{bowyer1985guide}: the fuel consumption rate $f_i$ ($\mathrm{mL/s}$) of the $i$-th vehicle is calculated as
	\begin{equation} \label{Eq:FuelModel}
	f_i  =  \begin{cases} 
	0.444 + 0.090 R_i v_i   +   [0.054 a_i^2 v_i] _{a_i>0},& \mathrm{if}\, R_i > 0,\\
	 0.444, & \mathrm{if} \, R_i  \le  0,
	\end{cases}
	\end{equation}
where $R_i = 0.333+0.00108 v_i^2 + 1.200 a_i$ ($a_i$ denotes the acceleration of vehicle $i$). 

{
\begin{table}[htb]
\footnotesize
	\begin{center}
		\caption{Parameter Setup of Distributed \method{DeeP-LCC} in the Large-Scale Simulation}\label{Tb:ParameterSetupLargeScale}
		\begin{threeparttable}
		\setlength{\tabcolsep}{5mm}{
		\begin{tabular}{cccccccc}
		\toprule
			Penetration rate & $m_i$ & $T_i$ & $w_v$ & $w_s$ & $w_u$ & $\lambda_{g_i}$ & $\lambda_{y_i}$ \\\hline
			 $5\%$ & $16,17,19,20,23$ & 800 & $2$ & $1$ & $0.2$ & $2$ & $10^4$\\
    $10\%$ & $7,8,9,10,11$ & 600 &  $1$ & $0.5$ & $0.1$ & $2$ & $10^4$\\
    $20\%$ & $3,4,5,6,7$ & 600 & $1$ & $0.5$ & $0.1$ & $2$ & $10^4$\\
			\bottomrule
		\end{tabular}}
		\end{threeparttable}
	\end{center}
	\vspace{-5mm}
\end{table}
}

\begin{table}[htb]
\footnotesize
	\begin{center}
		\caption{Velocity Profile of the Head Vehicle in the Large-Scale Simulation}\label{Tb:HeadVehicleTrajectory}
		\begin{threeparttable}
		\setlength{\tabcolsep}{7mm}{
		\begin{tabular}{cccccc}
		\toprule
			Time [$\mathrm{s}$] & $0-1$ & $1-4$ & $4-9$ & $9-150$ \\\hline
			 Acceleration [$\mathrm{m/s^2}$] & $-5$ & $0$ & $1$ & $0$\\
			\bottomrule
		\end{tabular}}
		\begin{tablenotes}
		\footnotesize
		\item[1] At the beginning of the simulation, there is one second time for initialization with $15\,\mathrm{m/s}$.
		\end{tablenotes}
		\end{threeparttable}
	\end{center}
	\vspace{-5mm}
\end{table}

\begin{figure*}[p]
	\centering
	\subfigure[All HDVs]
	{
	\includegraphics[width=8.6cm]{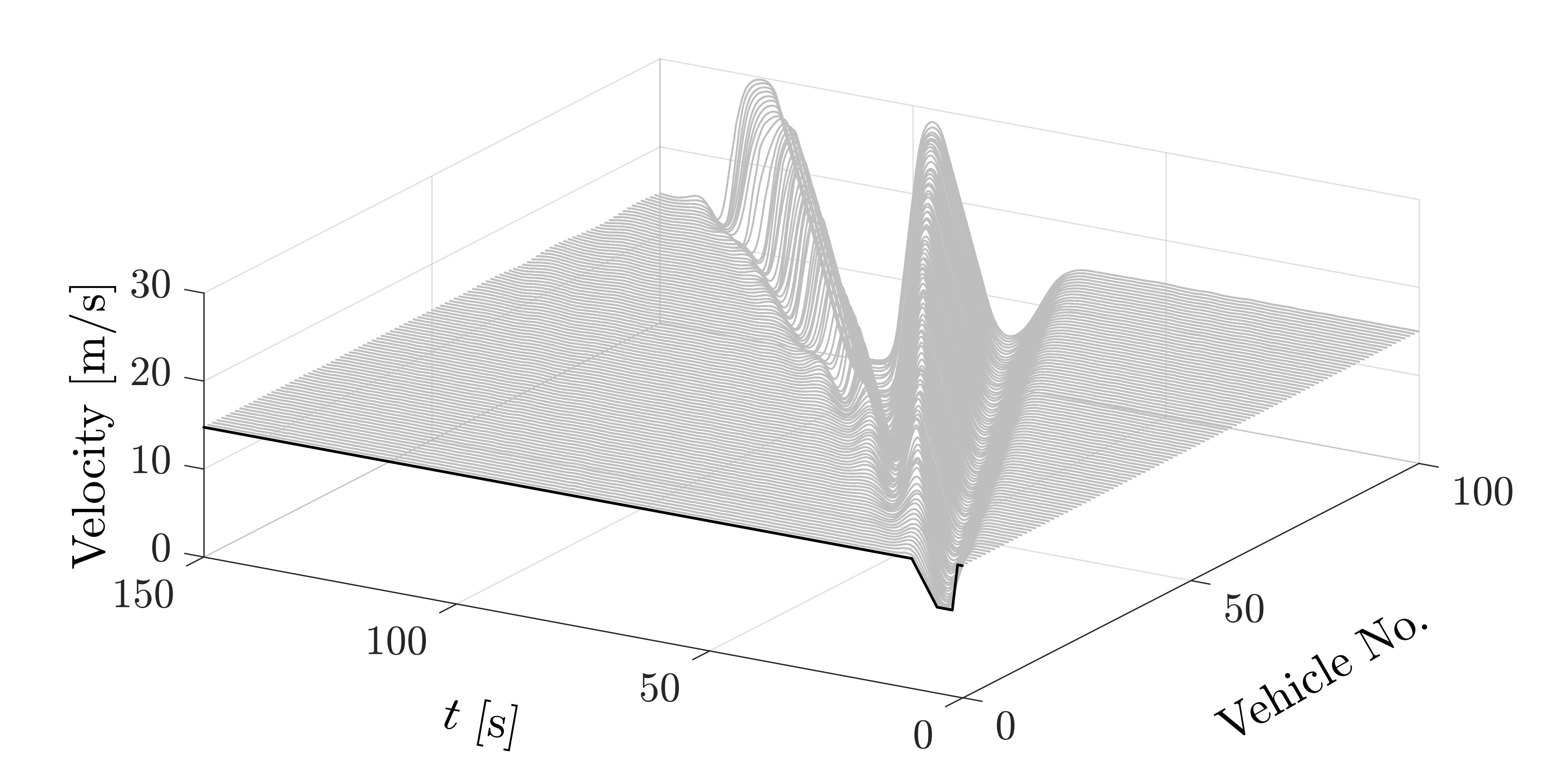}
	\includegraphics[width=6cm]{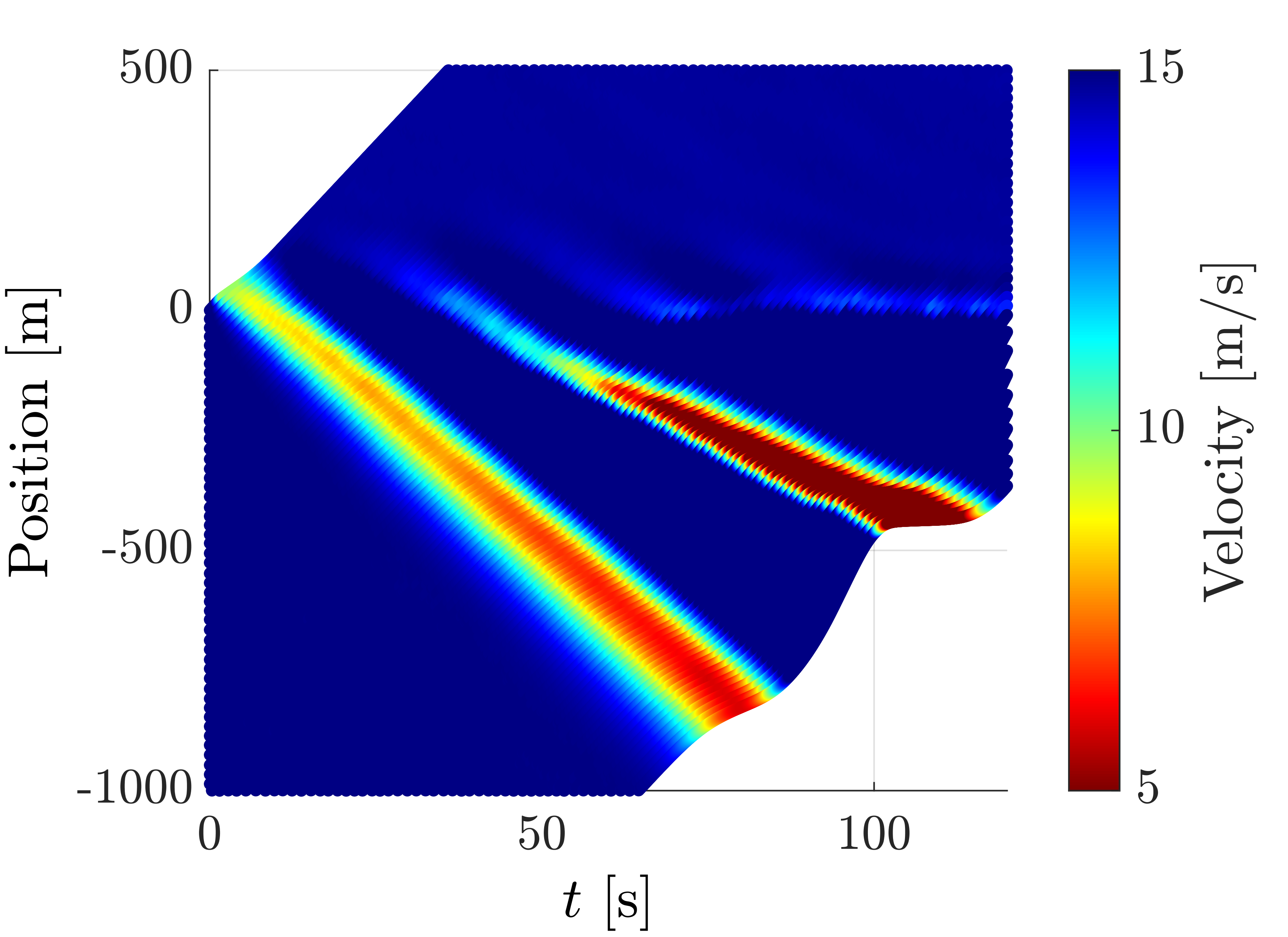}
	\label{Figure:LargeScale_HDVs}
	}\\
 \vspace{-2mm}
  \subfigure[Distributed \method{DeeP-LCC} with $5\%$ CAVs]
	{
	\includegraphics[width=8.6cm]{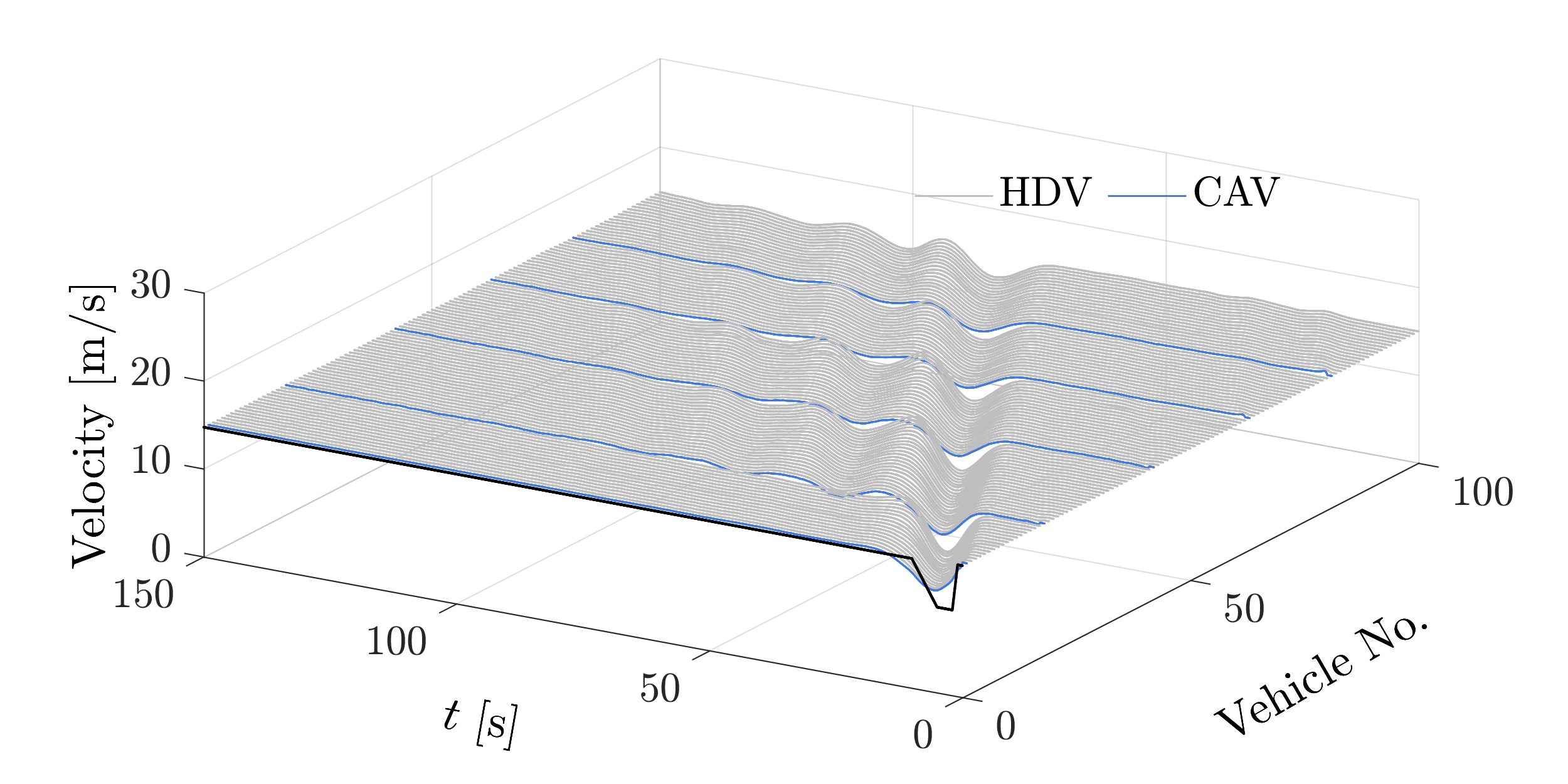}
	\includegraphics[width=6cm]{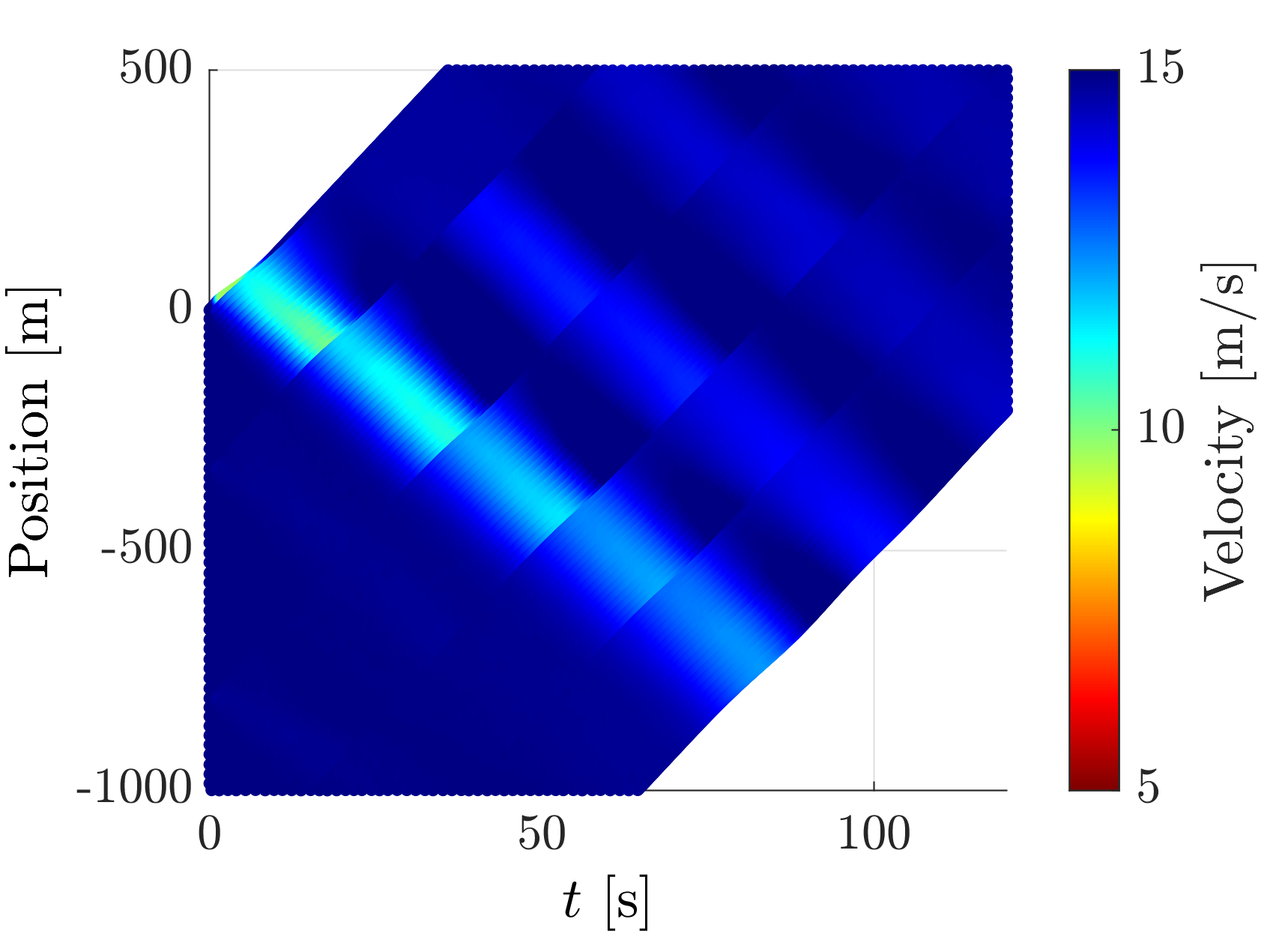}
	\label{Figure:LargeScale_CAVs_5}
	}\\
 \vspace{-2mm}
 \subfigure[Distributed \method{DeeP-LCC} with $10\%$ CAVs]
	{
	\includegraphics[width=8.6cm]{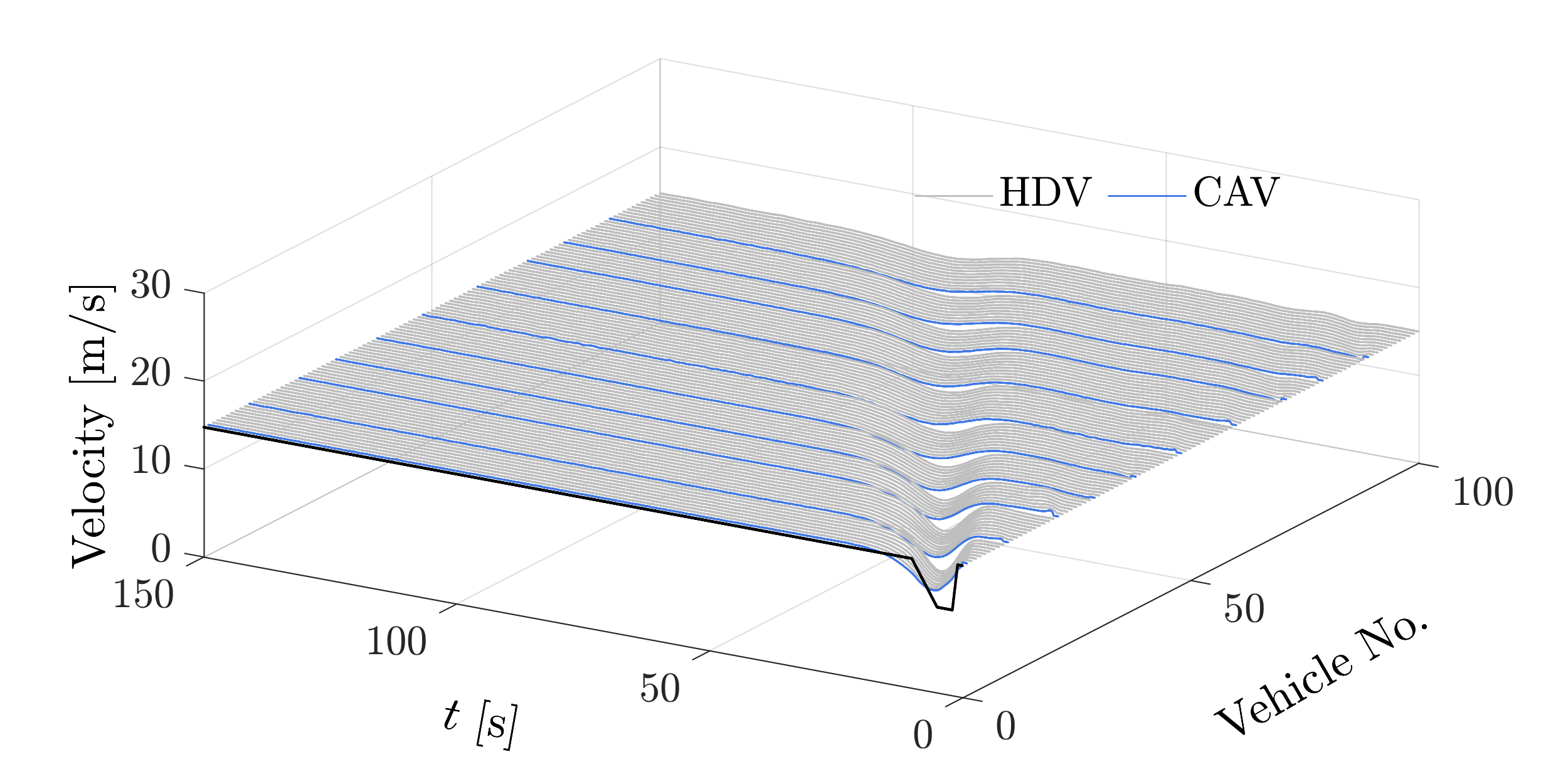}
	\includegraphics[width=6cm]{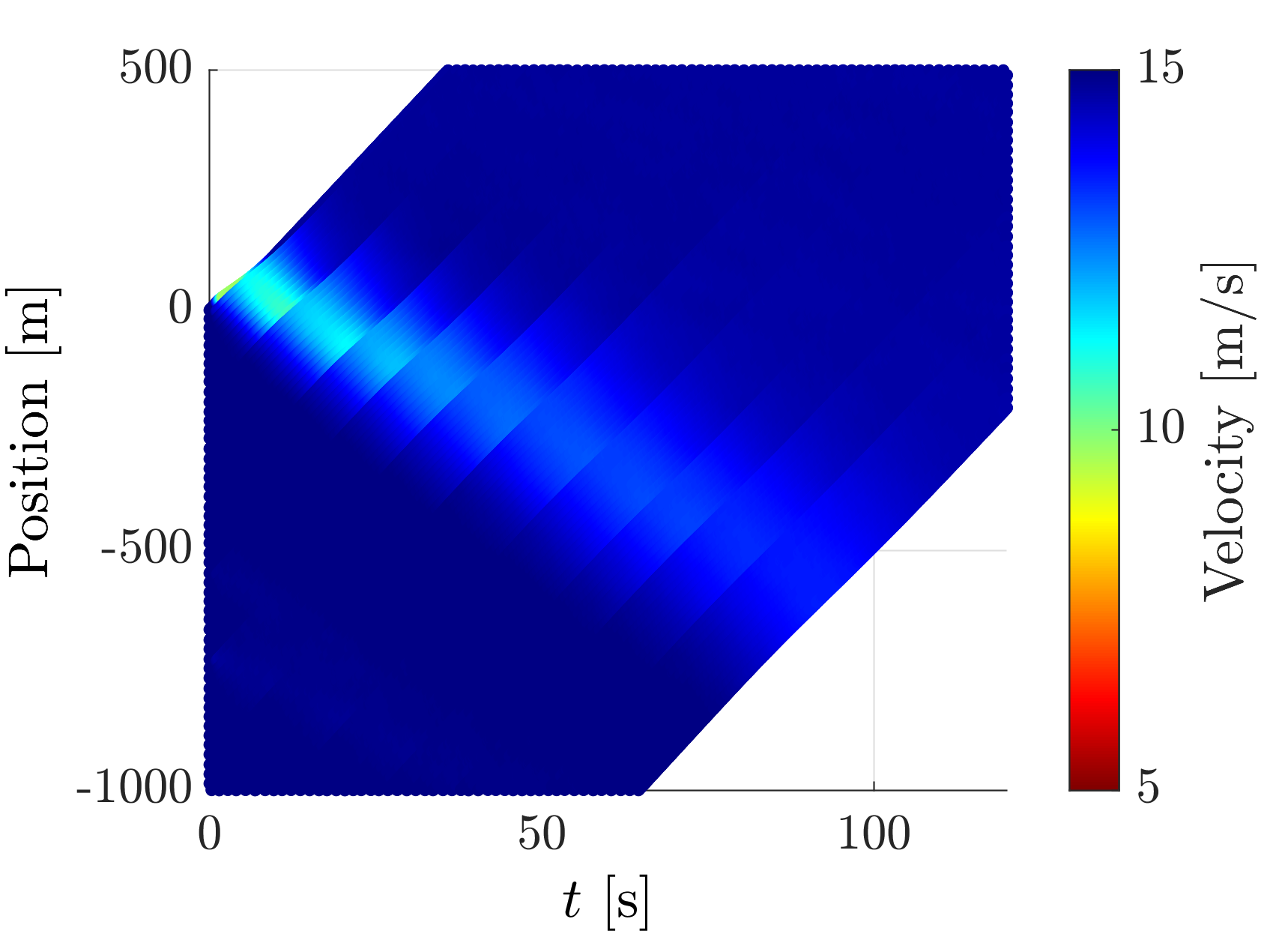}
	\label{Figure:LargeScale_CAVs_10}
	}\\
 \vspace{-2mm}
	\subfigure[Distributed \method{DeeP-LCC} with $20\%$ CAVs]
	{
	\includegraphics[width=8.6cm]{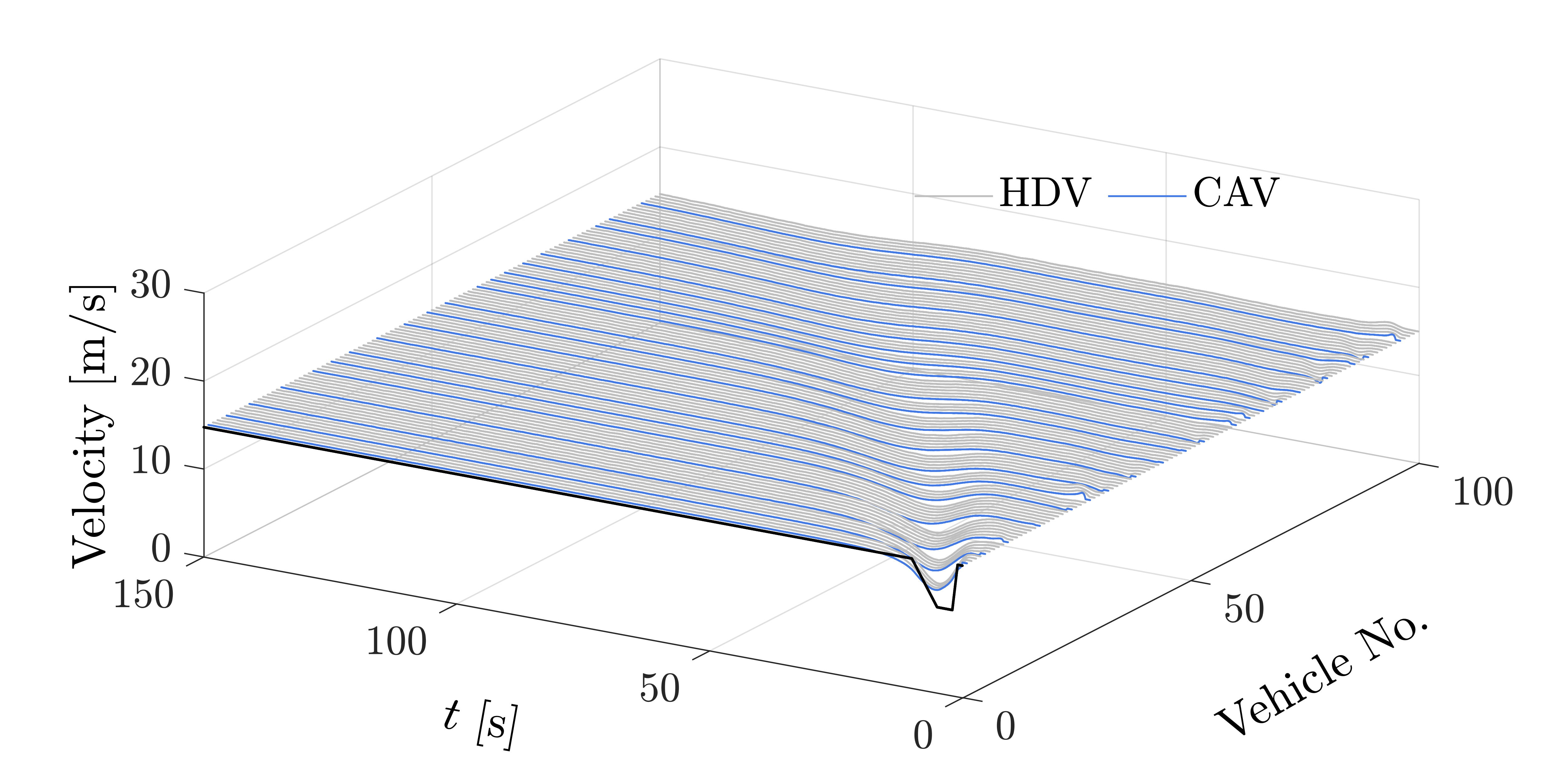}
	\includegraphics[width=6cm]{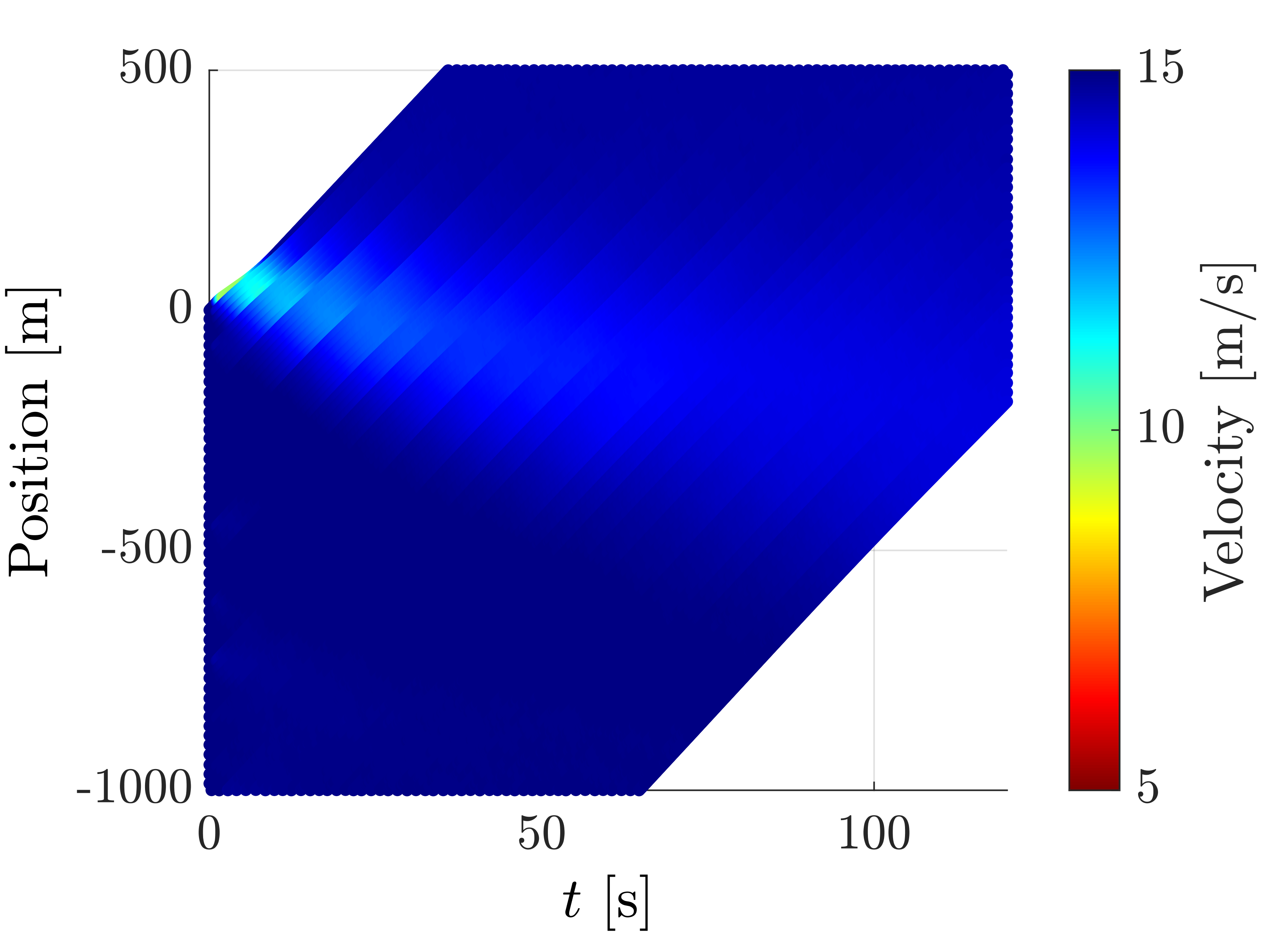}
	\label{Figure:LargeScale_CAVs}
	}
   \vspace{-2mm}
	\caption{Velocity profiles in large-scale experiments with $100$ vehicles, where the head vehicle is under a brake perturbation. In the right panel of each subfigure, the color denotes the vehicle velocity. (a) All the vehicles are HDVs. (b)-(d) The CAVs utilize the distributed \method{DeeP-LCC} controller, with different penetration rates. The black, blue, and gray profiles represents the head vehicle, the CAVs, and the HDVs, respectively.  }
	\label{Fig:LargeScaleSimulation}
\end{figure*}

As shown in Fig.~\ref{Figure:LargeScale_HDVs}, when all the vehicles are HDVs, the brake perturbation of the head vehicle causes a direct traffic wave propagating upstream, against the moving direction of the vehicles. Meanwhile, two additional waves are also observed (see the right panel of Fig.~\ref{Figure:LargeScale_HDVs}): one wave has slight velocity oscillations, while the other one shows the strongest oscillation amplitude. {In comparison, the traffic flow with a small proportion of CAVs equipped with distributed \method{DeeP-LCC} behaves quite smoothly in response to this brake perturbation, as can be clearly observed from Fig.~\ref{Figure:LargeScale_CAVs_5}-\ref{Figure:LargeScale_CAVs}. When there are only $5\%$ CAVs in mixed traffic flow, our method already enables the CAVs to apparently mitigate traffic waves. As the penetration rate grows up to $10\%$ or $20\%$, the traffic wave is rapidly dissipated, and most of the following vehicles only experience neglectable velocity oscillations.}

\begin{figure*}[t]
	\centering
	\subfigure[]
	{\includegraphics[width=7.5cm]{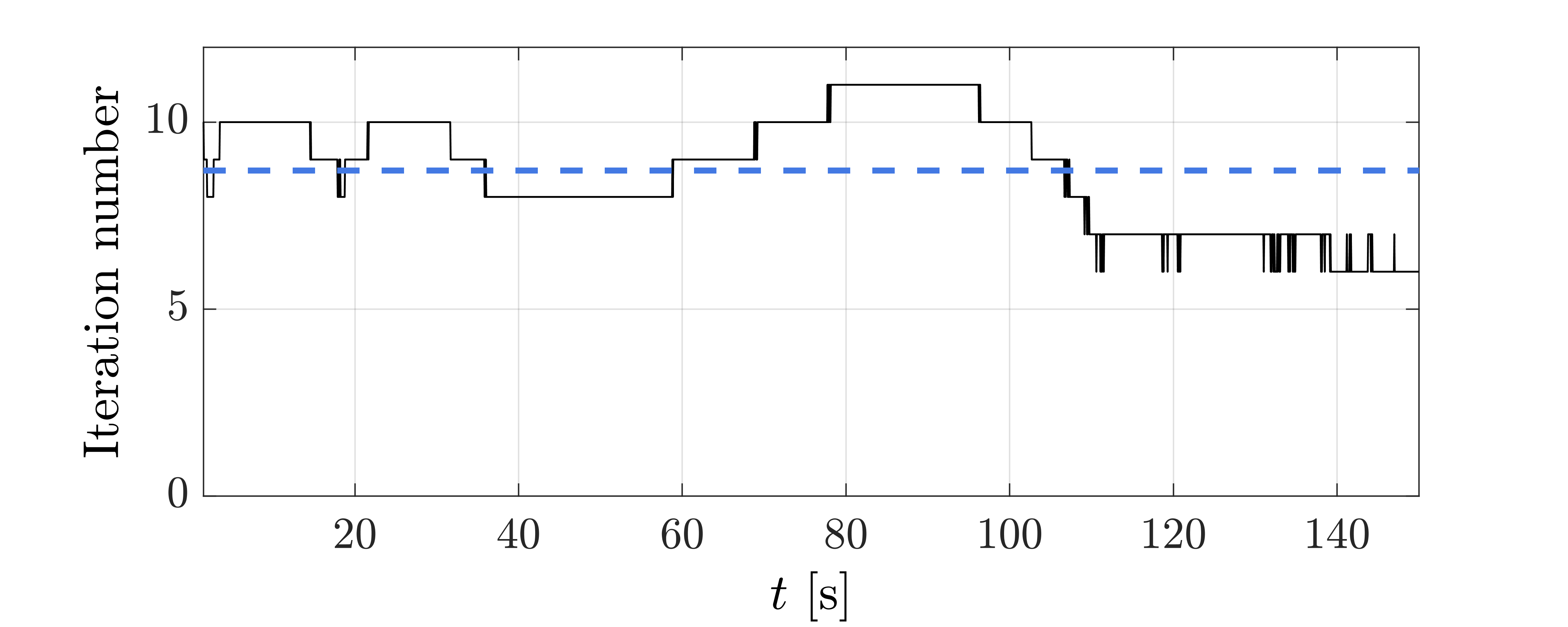}
	\label{Fig:LargeSinusoidPerturbation_HDV}}
	\subfigure[]
	{\includegraphics[width=7.5cm]{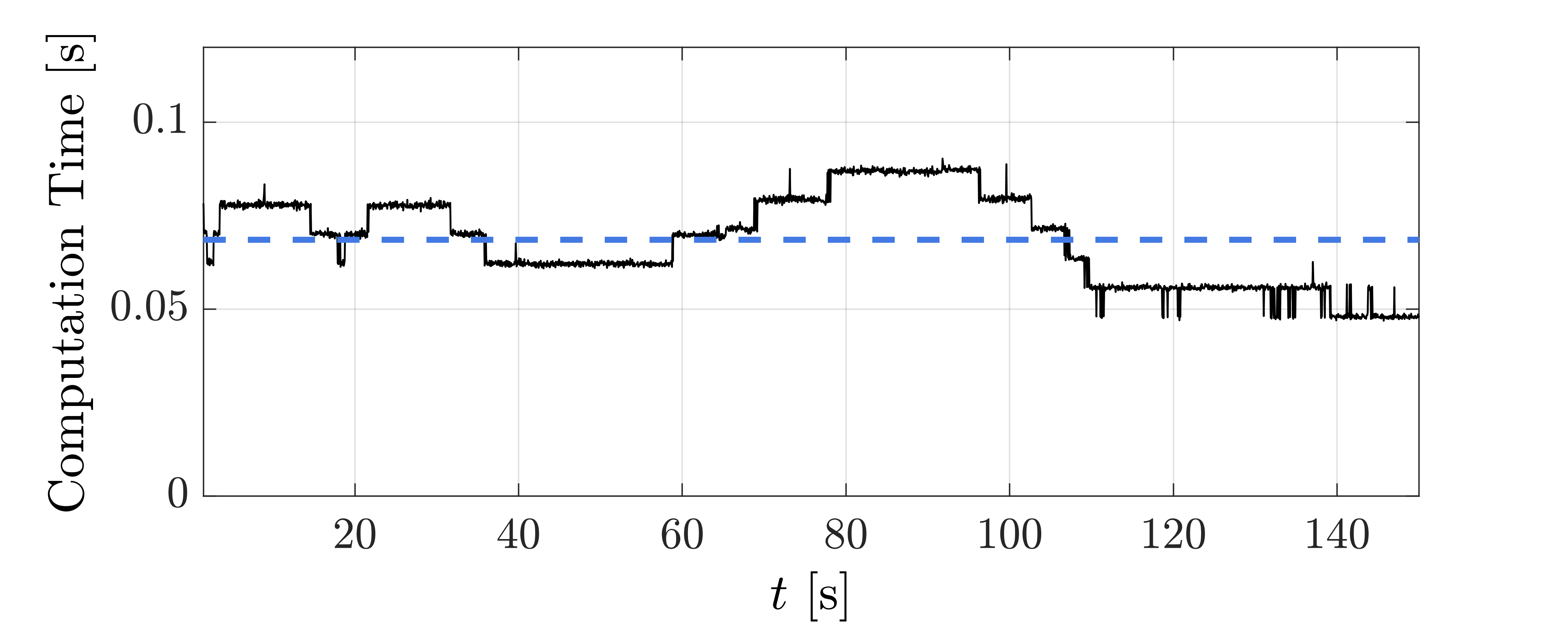}
	\label{Fig:LargeSinusoidPerturbation_CAV}}
	\caption{Iteration number and computation time of distributed \method{DeeP-LCC} at each time step under the penetration rate of $20\%$. The blue dashed lines represent the mean value throughout the simulations. (a) The average iteration number is $8.71$. (b) The average computation time is $0.069\,\mathrm{s}$.  This experiment is run in MATLAB 2021a with a CPU of Intel Core i7-11800H. The computation time could be further decreased via efficient software development.}
	\label{Fig:LargeScale_Statistics}
\end{figure*}

Fig.~\ref{Fig:LargeScale_Statistics} illustrates the iteration number and computation time of distributed \method{DeeP-LCC} in each time step under the penetration rate of $20\%$, and an average number of $8.70$ iterations is worth noting. In this large-scale mixed traffic system, it is intractable to solve centralized \method{DeeP-LCC} in real time, which would require a great number of data samples and become a large-scale quadratic programming problem. By contrast, our proposed distributed \method{DeeP-LCC} relies on local data of each subsystem, and with efficient ADMM design, it is capable of mitigating traffic perturbations in large-scale traffic flow. {Particularly, by employing the instantaneous fuel consumption model~\eqref{Eq:FuelModel}, a $31.84\%$, $32.34\%$, $32.53\%$ reduction of fuel consumption is achieved via distributed \method{DeeP-LCC} at the penetration rate of $5\%$, $10\%$, and $20\%$, respectively, compared to the case of pure HDVs. Indeed, we believe that the wave mitigation performance can be further improved by careful parameter tuning for distributed \method{DeeP-LCC} given different penetration rates and spatial formations of the CAVs in mixed traffic flow.}


\section{Conclusions}
\label{Sec:7}

In this paper, we have presented distributed \method{DeeP-LCC} for CAV cooperation in large-scale mixed traffic flow. We partition the entire mixed traffic system into multiple CF-LCC subsystems and establish a subsystem-based cooperative control formulation, where each CAV collects local data of its subsystem and utilizes a data-centric representation for predictive control. A tailored ADMM algorithm is designed for distributed implementation, which decomposes the coupling constraint and achieves computation and communication efficiency. Two different scales of traffic simulations confirm that distributed \method{DeeP-LCC} brings wave-dampening benefits with real-time computation performance. {One future direction is to verify the performance of distributed \method{DeeP-LCC} in real-world environments given the practical V2V/V2X communication capabilities.} Given the potential computation delay in each subsystem, another important topic is to develop asynchronous algorithms for distributed \method{DeeP-LCC} to support more robust application. {Finally, it would also be interesting to see the comparison between distributed \method{DeeP-LCC} with other data-driven methods for CAV control in mixed traffic.}

\section*{Acknowledgement}
The work of J. Wang, Q. Xu and K. Li is supported by National Key R\&D Program of China with 2021YFB1600402.
	
\appendix

\section{Proof of Theorem~\ref{Theorem:CostFunction}}
\label{Appendix:Proof}

The input and output definitions~\eqref{Eq:SystemOutput} and~\eqref{Eq:ControlInput} show that $u_i(t),y_i(t),\;i \in \mathbb{N}_1^n$ are indeed a row partition of $u(t),y(t)$. Thus, when Assumption~\ref{Assumption:SameTrajectory} holds, the pre-collected data $u_i^\mathrm{d},y_i^\mathrm{d},\;i \in \mathbb{N}_1^n$ are also a row partition of $u^\mathrm{d},y^\mathrm{d}$, we denote this partition pattern as $\mathcal{P}$. Given $T_i = T,\;i \in \mathbb{N}_1^n$, the data Hankel matrices $U_{i,\mathrm{p}},U_{i,\mathrm{f}}, Y_{i,\mathrm{p}},Y_{i,\mathrm{f}},\;i \in \mathbb{N}_1^n$ are a row partition of $U_{\mathrm{p}},U_{\mathrm{f}},Y_{\mathrm{p}},Y_{\mathrm{f}}$ respectively with pattern $\mathcal{P}$. Since a same past system trajectory is under consideration, $u_{i,\mathrm{ini}},y_{i,\mathrm{ini}},\;i \in \mathbb{N}_1^n$ are a row partition of $u_\mathrm{ini},y_\mathrm{ini}$ respectively with pattern $\mathcal{P}$. 

In addition, Assumption~\ref{Assumption:SameTrajectory} yields that for subsystem $1$, its external reference input is the same as that in the centralized formulation, and thus 
\begin{equation} \label{Eq:ProofExternalOne}
    E_{1,\mathrm{p}}=E_\mathrm{p},\,E_{1,\mathrm{f}}=E_\mathrm{f},\,\epsilon_{1,\mathrm{ini}}=\epsilon_\mathrm{ini}.
\end{equation}
Given the fact that the external reference input of the subsystems $2,3,\ldots,n$ is contained in the output of the subsystems $1,2,\ldots,n-1$ respectively, we have
\begin{equation} \label{Eq:ProofExternal}
    E_{i+1,\mathrm{p}}=K_{i}Y_{i,\mathrm{p}},\,E_{i+1,\mathrm{f}}=K_{i}Y_{i,\mathrm{f}},\,\epsilon_{i+1,\mathrm{ini}}=K_{i}y_{i,\mathrm{ini}},\;\;i\in \mathbb{N}_1^{n-1}.
\end{equation}

We first show that a feasible solution to Problem~\eqref{Eq:CooperativeDeePLCCLinearized} can be constructed based on $u^*,y^*$. Define $\overline{u}_i^*,\overline{y}_i^*,\;i\in \mathbb{N}_1^n$ as the $\mathcal{P}$-pattern row partition of $u^*,y^*$ respectively. Given the feasibility of $(g^*,u^*,y^*)$ for~\eqref{Eq:CentralizedDeePLCCLinearized} and the aforementioned partition properties, we have
\begin{subequations}
    \begin{align}
U_{i,\mathrm{p}}g^* &= u_{i,\mathrm{ini}},\,U_{i,\mathrm{f}}g^*= \overline{u}_i^*,\;\;i\in \mathbb{N}_1^n; \\
Y_{i,\mathrm{p}}g^* &= y_{i,\mathrm{ini}} ,\ \,Y_{i,\mathrm{f}}g^* = \overline{y}_i^*,\;\;i\in \mathbb{N}_1^n; \label{Eq:ProofOutput}\\
E_{1,\mathrm{p}}g^* &= \epsilon_{1,\mathrm{ini}},\, E_{1,\mathrm{f}}g^* = \hat{\epsilon},\;\;i\in \mathbb{N}_1^n.
\end{align}
\end{subequations}
Further, the following result is obtained by substituting~\eqref{Eq:ProofOutput} to~\eqref{Eq:ProofExternal} for all $i\in \mathbb{N}_1^{n-1}$
\begin{equation} \label{Eq:ProofExternalEquation}
E_{i+1,\mathrm{p}}g^*=K_iY_{i,\mathrm{p}}g^*=K_iy_{i,\mathrm{ini}}=\epsilon_{i+1,\mathrm{ini}},\quad 
E_{i+1,\mathrm{f}}g^* = K_iY_{i,\mathrm{f}}g^*=K_i\overline{y}_i^*.
\end{equation}
Based on~\eqref{Eq:ProofExternalOne}--\eqref{Eq:ProofExternalEquation}, we have
$$
\begin{bmatrix}
 U_{i,\mathrm{p}} \\ E_{i,\mathrm{p}}\\Y_{i,\mathrm{p}} \\ U_{i,\mathrm{f}} \\ E_{i,\mathrm{f}}\\ Y_{i,\mathrm{f}}
 \end{bmatrix}g^*=
 \begin{bmatrix}
 u_{i,\mathrm{ini}} \\ \epsilon_{i,\mathrm{ini}}\\ y_{i,\mathrm{ini}} \\ \overline{u}_i^* \\\epsilon_i \\ \overline{y}_i^*
 \end{bmatrix},
$$
with $\epsilon_1 = \hat{\epsilon}$ and $\epsilon_{i+1} = K_i\overline{y}_i^,\;i\in \mathbb{N}_1^{n-1}$. Given the consistency~\eqref{Eq:ConstraintConsistency} of input/output constraints between the entire system and the subsystems, it holds that $\overline{u}_i^*\in \mathcal{U}_i, \overline{y}_i^*\in \mathcal{Y}_i$. Therefore,
$(\overline{u}_i^*,\overline{y}_i^*,g^*),\;i\in \mathbb{N}_1^n$ is a feasible solution to~\eqref{Eq:CooperativeDeePLCCLinearized}. Thus, we have
\begin{equation*} \label{Eq:ProofVsmall}
   \sum_{i=1}^n V_i(u_i^*,y_i^*)  \leq \sum_{i=1}^n V_i(\overline{u}_i^*,\overline{y}_i^*).
\end{equation*}
Note that according to the definitions of the cost functions~\eqref{Eq:CostDefinition} and~\eqref{Eq:CostDefinitionSubsystem}, the following relationship is satisfied 
\begin{equation*} \label{Eq:ProofVequal}
    \sum_{i=1}^n V_i(\overline{u}_i^*,\overline{y}_i^*) = V(u^*,y^*). 
\end{equation*}
Hence, we have 
\begin{equation} \label{Eq:CostRelationship1}
    \sum_{i=1}^n V_i(u_i^*,y_i^*) \leq V(u^*,y^*).
\end{equation}

We proceed to show that a feasible solution to Problem~\eqref{Eq:CentralizedDeePLCCLinearized} can be constructed based on $u_i^*,y_i^*$. Precisely, define $\overline{u}^* = \begin{bmatrix} u_1^*,\ldots,u_n^* \end{bmatrix}$, $\overline{y}^* = \begin{bmatrix} y_1^*,\ldots,y_n^* \end{bmatrix}$. According to Assumption~\ref{Assumption:SameTrajectory} and the input and output definitions~\eqref{Eq:SystemOutput} and~\eqref{Eq:ControlInput}, it is straightforward to know that $(u_\mathrm{ini},\epsilon_\mathrm{ini},y_\mathrm{ini},\overline{u}^*,\hat{\epsilon},\overline{y}^*)$ is already a trajectory of the linearized mixed traffic system~\eqref{Eq:DT_TrafficModel}. By Lemma~\ref{Proposition:DeePCMixedTraffic}, there exists a $\overline{g}^*$ such that the equality constraint in Problem~\eqref{Eq:CentralizedDeePLCCLinearized} is satisfied. Given the consistency of $\overline{u}^* \in \mathcal{U},\overline{y}^*\in \mathcal{Y} \iff u_i^*\in \mathcal{U}_i,y_i^*\in \mathcal{Y}_i, \; i \in \mathbb{N}_1^n$, it can be derived that $(\overline{g}^*, \overline{u}^*,\overline{y}^*)$ is a feasible solution to~\eqref{Eq:CentralizedDeePLCCLinearized}. Hence, we have
$$
V(u^*,y^*) \leq V(\overline{u}^*,\overline{y}^*).
$$
Meanwhile, we have
$$
V(\overline{u}^*,\overline{y}^*) = \sum_{i=1}^n V_i(u_i^*,y_i^*),
$$
and thus it holds that
\begin{equation} \label{Eq:CostRelationship2}
    V(u^*,y^*) \leq \sum_{i=1}^n V_i(u_i^*,y_i^*).
\end{equation}
Combining~\eqref{Eq:CostRelationship1} and~\eqref{Eq:CostRelationship2} leads to the result in~\eqref{Eq:CostRelationship}.

\section{Stopping Criterion of distributed \method{DeeP-LCC}}
\label{Appendix:Stopping}

Algorithm~\ref{Alg:DistributedDeePLCC} iterates until $300$ rounds or the following stopping criteria is satisfied  $(i=1,2,3,4)$
\begin{equation} \label{Eq:StoppingCriteria}
     r^{(i)}_{\mathrm{pri}}  \leq  \delta^{(i)}_{\mathrm{pri}} \ \mathrm{and} \  r^{(i)}_{\mathrm{dual}}  \leq  \delta^{(i)}_{\mathrm{dual}},
\end{equation}
where $r^{(i)}_{\mathrm{pri}},r^{(i)}_{\mathrm{dual}}$ denote the summarized two-norm value of primal and dual residuals respectively, defined as
$$
\begin{aligned} 
    r^{(1)}_{\mathrm{pri}}  =  &\sum_{i=1}^n   \left\| g_i^+  -   z_i^+ \right\|_2 , \ r^{(2)}_{\mathrm{pri}}   =   \sum_{i=1}^{n-1}   \left\| E_{i+1,\mathrm{f}}g_{i+1}^+    -    K_i Y_{i,\mathrm{f}}z_i^+ \right\|_2 , \\
    r^{(3)}_{\mathrm{pri}}  =  &\sum_{i=1}^n  \left\| \tilde{s}_i^+  -  P_i Y_{i,\mathrm{f}} g_i^+ \right\|_2 , \ r^{(4)}_{\mathrm{pri}}   =   \sum_{i=1}^n   \left\| u_i^+   -    U_{i,\mathrm{f}}g_i \right\|_2 ; \\
    r^{(1)}_{\mathrm{dual}}  =  & \sum_{i=1}^n  \rho  \left\| z_i^+  -  z_i \right\|_2  , \  r^{(2)}_{\mathrm{dual}}   =   \sum_{i=1}^{n-1}   \rho   \left\| E_{i+1,\mathrm{f}}^\top K_i Y_{i,\mathrm{f}}( z_i^+    -    z_i) \right\|_2  , \\
    r^{(3)}_{\mathrm{dual}}  =  & \sum_{i=1}^n  \rho   \left\| P_i^\top Y_{i,\mathrm{f}}^\top (\tilde{s}_i^+    -  \tilde{s}_i) \right\|_2  , \ 
     r^{(4)}_{\mathrm{dual}}  =  \sum_{i=1}^n  \rho  \left\| U_{i,\mathrm{f}}^\top (u_i^+   -   u_i)  \right\|_2 ,
\end{aligned}
$$
which corresponds to the four equality constraints~\eqref{Eq:ADMMFinalProblemGZequal}--\eqref{Eq:ADMMFinalProblemU}. In~\eqref{Eq:StoppingCriteria}, $\delta^{(i)}_{\mathrm{pri}}$ and $\delta^{(i)}_{\mathrm{dual}}$ denote the feasibility tolerances. Given a series of  equality constraints $Ax_i=By_i$  with dual variables~$\kappa_i$ $(i\in \mathcal{S})$, which is in the general form of~\eqref{Eq:ADMMFinalProblemGZequal}--\eqref{Eq:ADMMFinalProblemU}, $\delta_{\mathrm{pri}}$ and $\delta_{\mathrm{dual}}$  are chosen by the following rule~\cite{boyd2011distributed}
$$
\delta_{\mathrm{pri}}  = \sum_{i\in \mathcal{S}} \sqrt{k}\delta_\mathrm{abs} + \delta_\mathrm{rel}\max\{\left\|Ax_i^+\right\|_2,\left\|By_i^+\right\|_2\} ,\quad 
\delta_{\mathrm{dual}}  = \sum_{i\in \mathcal{S}} \sqrt{l}\delta_\mathrm{abs} +\delta_\mathrm{rel}\left\| A^\top \kappa_i^+ \right\|_2,
$$
where $\delta_\mathrm{abs},\delta_\mathrm{rel}$ denote an absolute and relative tolerance respectively, and $k,l$ represent the size of the corresponding $\ell_2$ norm in each formula.

	
\bibliography{mybibfile}

\end{document}